\documentclass[journal]{IEEEtran}

\ifCLASSINFOpdf
\usepackage{amsmath,graphicx,url,times,booktabs, tabularx}
\usepackage{threeparttable}
\usepackage{multirow}
\usepackage{gensymb}
\usepackage{subfigure}
\usepackage{pifont}

\usepackage{amssymb}
\usepackage{url}
\def\UrlBreaks{\do\A\do\B\do\C\do\D\do\E\do\F\do\G\do\H\do\I\do\J

\do\K\do\L\do\M\do\N\do\O\do\P\do\Q\do\R\do\S\do\T\do\U\do\V

\do\W\do\X\do\Y\do\Z\do\[\do\\\do\]\do\^\do\_\do\`\do\a\do\b

\do\c\do\d\do\e\do\f\do\g\do\h\do\i\do\j\do\k\do\l\do\m\do\n

\do\o\do\p\do\q\do\r\do\s\do\t\do\u\do\v\do\w\do\x\do\y\do\z

\do\.\do\@\do\\\do\/\do\!\do\_\do\ \do\;\do\>\do\]\do\)\do\,

\do\?\do\'\do+\do\=\do\#}

\else
\fi
\usepackage{array}
\begin{document}
%
\title{Dynamic Kernel Convolution Network 

with Scene-dedicate Training for

Sound Event Localization and Detection}
%
%
%

\author{Siwei Huang,~
Jianfeng Chen,~\IEEEmembership{Senior Member,~IEEE,}~Jisheng Bai,~\IEEEmembership{Student~Member,~IEEE,}~Yafei Jia, ~Dongzhe Zhang

\thanks{This work is supported by School of Marine Science and Technology, Northwestern Polytechnical University-LianFeng Acoustic Joint Laboratory Innovation Fund Project (No. 
 LFXS-JLESS-KT20220401). \textit{(Corresponding author: Jianfeng Chen.)} }
\thanks{S. Huang, J. Bai, Y. Jia and J. chen are with the Joint Laboratory of Environmental Sound Sensing, School of Marine Science and Technology, Northwestern Polytechnical University, Xi’an, China, and also with the LianFeng Acoustic Technologies Co., Ltd. Xi'an, China. (e-mail:hsw838866721@mail.nwpu.edu.cn; baijs@mail.nwpu.edu.cn; 
 jyf2020260709@mail.nwpu.edu.cn; chenjf@nwpu.edu.cn)}

}

%
%

\markboth{Journal of \LaTeX\ Class Files,~Vol.~14, No.~8, August~2015}%
{Shell \MakeLowercase{\textit{et al.}}: Bare Demo of IEEEtran.cls for IEEE Journals}
%



\maketitle

\begin{abstract}

DNN-based methods have shown high performance in sound event localization and detection(SELD). While in real spatial sound scenes, reverberation and the imbalanced presence of various sound events increase the complexity of the SELD task. In this paper, we propose an effective SELD system in real spatial scenes.
In our approach, a dynamic kernel convolution module is introduced after the convolution blocks to adaptively model the channel-wise features with different receptive fields. 
Secondly, we incorporate the  SELDnet and EINv2 framework into the proposed SELD system with multi-track ACCDOA. 
Moreover, two scene-dedicated strategies are introduced into the training stage to improve the generalization of the system in realistic spatial sound scenes. 
Finally, we apply data augmentation methods to extend the dataset using channel rotation, spatial data synthesis. Four joint metrics are used to evaluate the performance of the SELD system on the Sony-TAu Realistic Spatial Soundscapes 2022 dataset.
Experimental results show that the proposed systems outperform the fixed-kernel convolution SELD systems. In addition, the proposed system achieved an SELD score of 0.348 in the DCASE SELD task and surpassed the SOTA methods.
\end{abstract}


%
\IEEEpeerreviewmaketitle

\section{Introduction}
\label{sec:intro}
%
%
%
%
\IEEEPARstart{S}{ound} event localization and detection (SELD) is a task that involves sound onset, offset detection (SED), and the corresponding direction of arrival (DOA) estimation  using multi-channel spatial audios. A SELD system usually use time-frequency representations and magnitude/phase differences to recognize sound classes and directions.
SELD systems can be potentially used in many applications, such as surveillance systems \cite{Surveillance}, wildlife monitoring \cite{stowell2016bird}, virtual reality techniques \cite{blauert1997spatial}, and outdoor navigation. Nowadays, SELD task has been widely participated since a series of competitions are successfully held, for example, the Detection and Classification of
Acoustic Scenes and Events (DCASE) challenge    \cite{Adavanne2018_JSTSP} and IEEE  ICASSP Grand Challenge - L3DAS \cite{l3das}. Most SELD systems are implemented with multichannel audio files from a microphone array of two different formats in current use: first-order ambisonic (FOA) and multichannel microphone array (MIC). Recent challenges have been introduing more complex conditions into the SELD task, such as interference, overlapping sources, moving sources, and reverberation, to make the research of SELD has been heading further.

\begin{figure}[t]
    \centering
    \includegraphics[width=8.5cm]{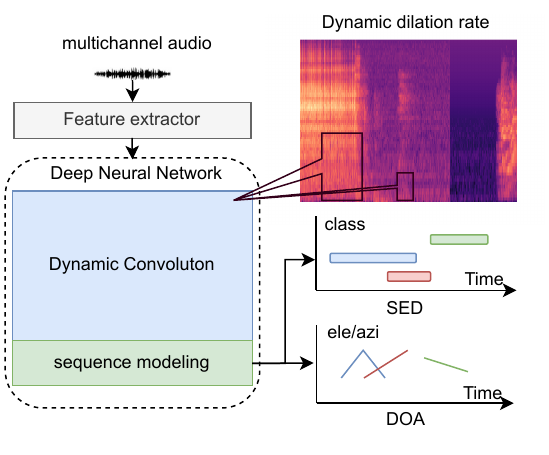}
    \caption{The sketch of Sound Event Localization and Detection(SELD) system.}
    \label{fig:feature}
\end{figure}

SELD research has predominantly relied on simulated data for training and evaluation. By utilizing real audio clips \cite{fonseca2022FSD50K} and pre-recorded spatial responses \cite{politis2020dataset}, it is theoretically possible to synthesize sound event datasets with annotated positional information for training and testing purposes. In 2022, the introduction of real-world SELD datasets with real-time spatial annotations by  Politis et al. \cite{star} has brought forth several challenges. Sound events exhibit increasing diversity, with some noise bands being wide and long-lasting, while others emit narrowband signals in short durations. The multi-scale nature of sound features poses challenges for unified modeling by SELD models. Additionally, the high cost of spatial annotation in real environments limits the size of real-world SELD datasets, resulting in extreme data class imbalances and instances where certain sound categories may occur only briefly or not at all in specific scenes. Models often struggle to achieve sufficient learning performance with such datasets.

In this paper, we introduce a dynamic kernel (DK) \cite{DK} convolutional module into the SELD system.  This convolutional module utilizes convolutional kernels with varying dilation rates to extract features from acoustic spectrograms. The extracted features are then aggregated using a channel attention mechanism \cite{SK}, which combines different scale feature maps based on their weights. This approach enables adaptive selection of scales during the inference process, facilitating effective detection of multi-scale sound events. By aliasing DK module and Conformer block, we propose three modified SELD networks based on the SOTA architecture for SELD tasks in real spatial scenes. Furthermore, we propose two training strategies to maximize the benefits of simulated datasets and address generalization issues on real datasets, which is named scene-dedicate training.
 The SELD system is evaluated using the Sony-TAu Realistic Spatial Soundscapes 2022 dataset (STARSS22) \cite{star}. Joint metrics and the aggregated SELD score are used to evaluate the performance of the SELD system. 
Experimental results reveal that the proposed  SELD network has outperformed the baseline SELD system with Fixed-size convolution kernels.
In addition, an ensemble system of proposed methods in the early stage submitted to the 2022 DCASE SELD task has achieved an overwhelming performance for microphone array data. Furthermore, the ablation study proved the effectiveness of modules, training strategies, and augmentations.

Our main contributions can be summarized as follows:
\begin{enumerate}
    \item[1)]Incorporate a dynamic convolution module into SELDnet and EINv2 framework for SELD task, to capture channel-wise features with large and dynamic receptive fields. 
    \item[2)]Design scenes-dedicated training methods. Improve the SELD performance evaluated on the specific scenes when the model is exposed to different sound fields.
    \item[3)]The proposed SELD system achieves SOTA performance and ranks first on the MIC dataset based on DCASE 2022.

\end{enumerate}

The rest paper is organized as follows:  Section \ref{sec:work} introduces the related techniques of SELD. Section   \ref{sec:net} describes our proposed methods, including network design, scene-dedicate training, and data augmentation. Section \ref{sec:exp} presents experimental settings and a comparison of other SELD systems on the STARSS22 blind test set. Based on the development set of STARSS22. We further show experimental results and discussions of proposed methods with a set of ablation studies. 
Finally, we conclude our work in Section \ref{sec:conclusion}.

\section{Related works}
\label{sec:work}
\subsection{Existing SELD methods}
In the past few years, there have been many major developments in the structures of SELD systems. In 2018, Adavanne
et al. \cite{Adavanne2018_JSTSP} proposed a polyphonic SELD system with a unified convolutional recurrent neural network, named SELDnet. This system extracts features of multichannel audios, use CNNs to locally model the features and RNNs to achieve time series prediction. And the SED and DOA results are outputted through two separate fully-connect branches. In 2019, Cao et al. \cite{cao2019polyphonic} proposed a two-stage strategy that separately trains SED and DOA models. In 2020, Wang et al. \cite{seld_TDNNF} adopted ResNet \cite{he2016deep} architectures as the high-level feature representation module and factorized time delay neural network (TDNNF) \cite{Povey2018SemiOrthogonalLM} structure for temporal context modeling. In 2020, Cao et al. \cite{cao2020event} proposed an Event-Independent (EIN) Network for polyphonic SELD, which uses soft parameter sharing between SED and DOA branch and outputs track-wise predictions via permutation invariant training(PIT). Shimada et al. \cite{accdoa} proposed an Activity-Coupled Cartesian Direction of Arrival (ACCDOA) representation that combines SED and DOA loss into one regressing function. EINv2 \cite{9413473} is an improved work of  \cite{cao2020event} that replaces RNN with multi-head self-attention (MHSA) mechanism \cite{vaswani2017attention}. In 2021,
Lee et al. \cite{app12105075} deemed SELD as a multi-modal task and improved EINv2 by adding cross-modal attention between SED and DOA branches. Shimada et al. \cite{Shimada2022} extend their ACCDOA representation to multi-track ACCDOA and adapted auxiliary duplicating permutation invariant (ADPIT) to make track-wise prediction available.

\subsection{Multi-scale representation}
Polyphonic sound, which is captured as an aliasing signal and contorts the representation of features from that of any specific target, has always been an important fact for the degration of SELD perfermance in SELD. It is crucial to extract high-resolution acoustic features that represent sound events in different scales. Shimada et al.  \cite{Shimada_SONY_task3_report} proposed RD3Net based on a densely connected mutilated DenseNet (D3Net), incorporating multi-resolution learning with an exponentially growing receptive field. And the RD3Net-based SELD system has achieved the best performance in the 2021 DCASE SELD task.
Zheng et al. \cite{SK,Zheng2021} added an event-aware module into their CRNN models. The selective kernel improves the performance of detecting class-wise sound events in different scales. 
 Nguyen et al. \cite{nguyen2022salsa} proposed a novel feature that stacks multichannel log-spectrograms and the normalized principal eigenvector of the spatial covariance matrix at each corresponding time-frequency bin, called SALSA. And their improved work \cite{Nguyen2022} replaces the principal eigenvector with phase differences between microphones, which costs less computing resources.  In 2022, multi-channel data recorded in real spatial scenes were introduced in the DCASE SELD task, Xie et al. \cite{Xie_UESTC_task3_report}  incorporated convolutional block attention module (CBAM) in their ensemble system, which shows a robust SELD performence in multiple interference and noise scenarios. Ko et al. \cite{Ko_KAIST_task3_report} submitted a SELD system with Squeeze-and-Excitation (SE) network, which handles multiple-dimension feature capturing. Xie et al. \cite{Xie_XJU_task3_report} design a CRNN architecture using the Time-Frequency attention (TTA) and the criss-cross Attention (CCA). All systems have adopted various multi-scale modules or attention mechanisms and achieved great performance in the 2022 DCASE SELD task, revealing their effectiveness in complex real spatial scenes.

\section{Proposed method}
\label{sec:net}
In this section, we introduce the dynamic kernel convolution module and the Conformer encoder in the SELD structures. We further introduce scene-dedicate training and a series of data augmentation methods in the model training procedure.

\subsection{Feature extraction}
\label{ssec:feature}
In the SELD task, the input feature is generally split into two major parts: the SED feature and the DOA feature. The SED feature denotes the time-frequency representation of each channel, and the DOA feature denotes the spatial representation like phase information and magnitudes in different directions. As mentioned in Section \ref{sec:intro}, features like linear spectrogram, log-Mel spectrogram, and MFCC \cite{MFCC} of each channel are generally extracted as acoustic representation. And the magnitude difference of each channel, the phase of the STFT matrix, and the time delay estimation of each channel are computed as  spatial representations.  And GCC-PHAT for MIC data estimates the time delay of the dominant sound source signal arriving at different receivers \cite{Adavanne2018_JSTSP}.

 We extract the log-Mel spectrogram of each channel as SED feature and the frequency-normalized inter-phase difference(NIPD) \cite{araki2007underdetermined} in SALSA-Lite  as DOA feature, named SALSA-Mel. The Mel-filter bank consists of filters with narrower bandwidths as the frequency decrease, and the Mel-scale frequency matches the sensitivity of human auditory perception when frequency changes.
For a multichannel recording with M channels, the log-Mel spectrogram is extracted via:
\begin{align}
Spec(t, bin) = 20\log(\sum_{f}^{F}H_{m}(f,bin)|X_{1:M}(t,f)|),
\end{align}
where $X_{i}(t,f)$ is the short-time Fourier transform (STFT) matrix of the $i^{th}$ channel.$H_{m}(f,bin)$ is the frequency weights matrix of Mel filters, $bin$ defines a filter bin in Mel filters, the NIPD in SALSA-Lite is computed via:
\begin{align}
\Lambda(t,f) = -c(2{\pi}f)^{-1}\arg({X_{1}^{*}}(t,f){X_{2:M}}(t,f)),
\end{align}
where $\cdot^{*}$ is its conjugate transpose matrix. As the original SALSA-Lite is applied to the log-linear spectrogram, we multiply it with Mel filters so that both features are in the same shape. It is noteworthy that,  the weighted summation of phase difference creates a new pattern and enlarges the focus area of the low-frequency phase i.e. the potent part without spatial aliasing. SALSA-Mel is utilized as the summary of the proposed SELD feature from the original SALSA-Lite for the next Sections.

\begin{figure}[t]
    \centering
    \includegraphics[width=6cm]{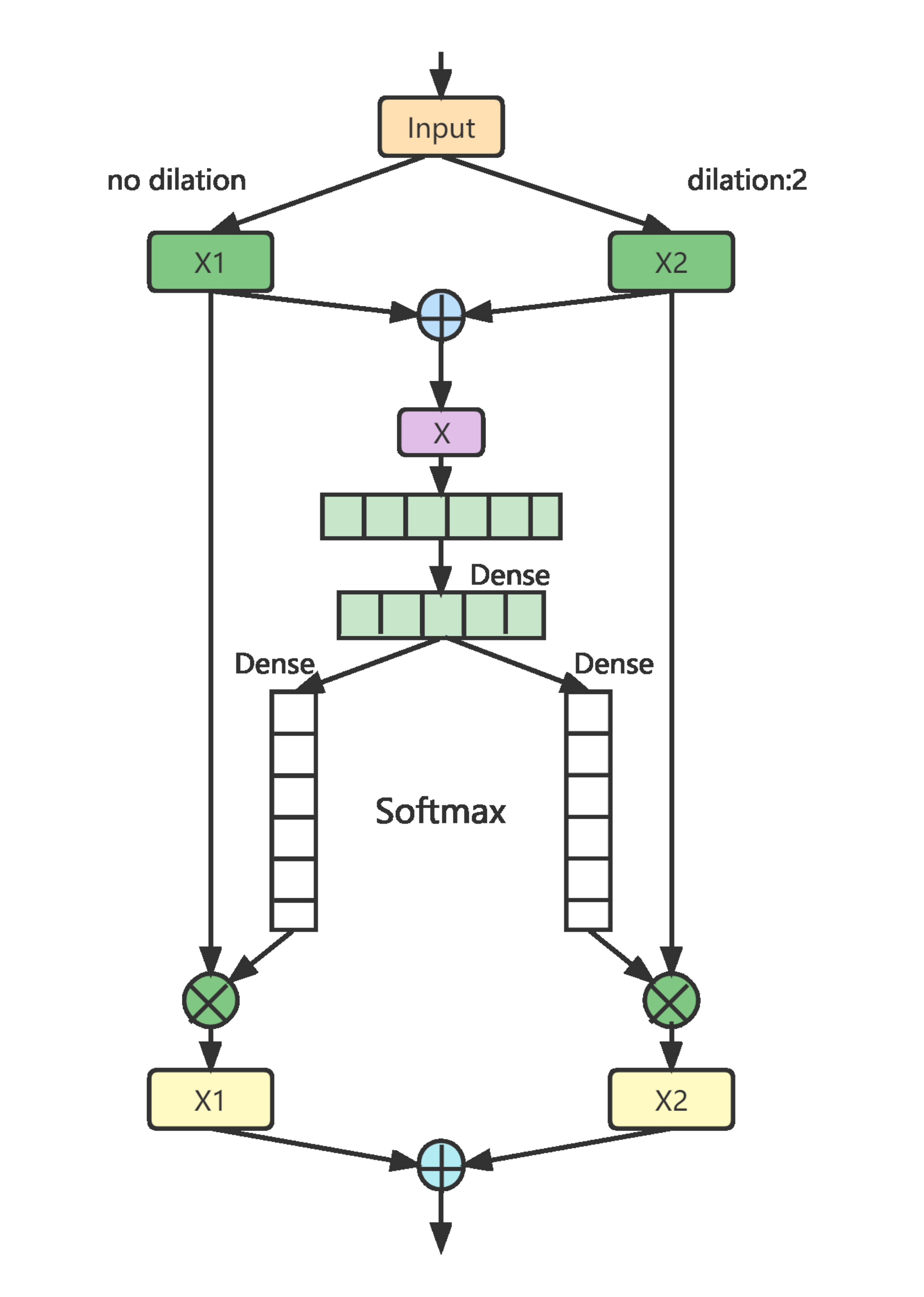}
    \caption{The architecture of DK convolution module. $\bigoplus$ denotes element-wise addition.$\bigotimes$ denotes element-wise multiplication.}
    \label{fig:dk}
\end{figure}

 \begin{figure*}[h]
    \centering
    \includegraphics[width=18cm]{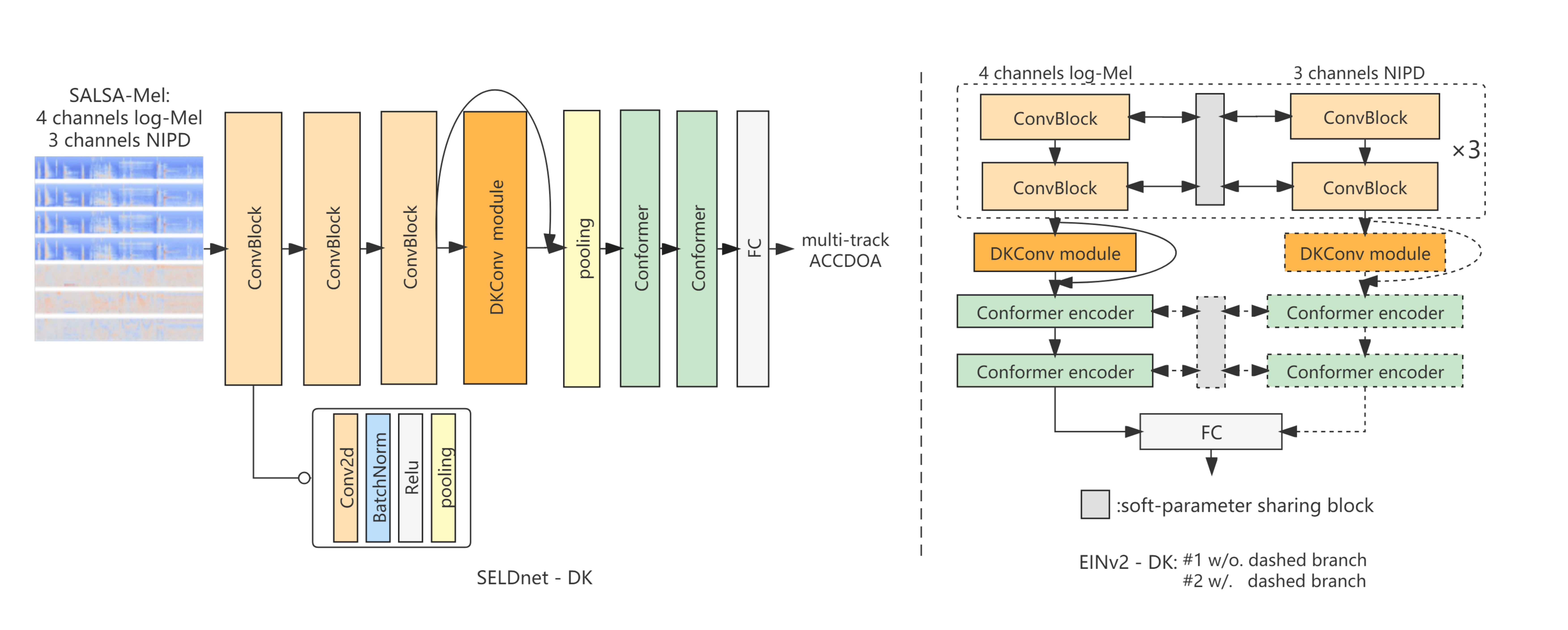}
    \caption{The constructions of three models. For SELDnet-DK, the DK convolution module and the Conformer encoder are incorporated into SELDnet, and the EINv2 framework is modified into two structures. Each of them shares single or two branches of weight respectively,  shortened to EINv2-DK\#1 and EINv2-DK\#2.  soft-parameters sharing block denotes the cross-stitch block  \cite{misra2016cross} in the EINv2 framework.}
    \label{fig:net}
\end{figure*}

\subsection{Dynamic kernel convolution based SELD network}
\label{ssec:DK}
In this paper, we incorporate dynamic kernel(DK) convolution into the SELD model and propose three  structures based on CNN-Conformer and EINv2 framework. Giving an overview of the model, as shown in Fig \ref{fig:net}, two DK convolution modules followed by the residual connection are applied after the final convolutional blocks, and the high-dimensional embeddings are fed into the Conformer encoder blocks. 

The architecture of the DK convolution module is illustrated in Fig \ref{fig:dk}. The DK convolution module is a dynamic channel selection mechanism based on softmax attention \cite{selective}, which consists of two branches of dilation convolution with the same kernel size and various dilation rates. By summing up the features selected from different branches, the module can adaptively capture features between long-term and short-term contexts.
Firstly, input features are convolved with  kernels of the same size and different dilation rates. We fuse the output features from different branches via an element-wise summation:
\begin{flalign}
\mit{X} =X_{1} + X_{2}.
\end{flalign}
Secondly, the channel embeddings $\mit{Z} \in{\Bbb{R}^C}$ are gathered using time and frequency average pooling. Specifically for the c-th element,  the channel-wise information is shrunk from the summary of the input feature map with time and frequency information:
\begin{flalign}
\mit{Z_{c}}=\frac{1} {T\times F}   \sum_{i=1}^T \sum_{j=1} ^F  \mit{X_c}(i,j).
\end{flalign}
Given a reduction rate $\mit{r}$, channel information is further reduced to a compact feature
$\mit{z} \in{\Bbb{R}^{C/r}}$,which is achieved using a fully connected(FC) layer and ReLU function. After that, channel-weight $\mit{s}$ for the ith branch is obtained via:   
\begin{flalign}
\mit{s_{i}}=\tau (\mit{W_{i} ^T} \mit(z) + b_{i})=\tau(\mit{W_{i} ^T  \mu(V_{i}^T Z + n_{i})}+b_{i}),
\end{flalign}
where $ \mit{V_{i}} \in{\Bbb{R}^{C/r \times C}}$, $\mit{W_{i}}   \in{\Bbb{R}^{C \times C/r}}$  
are weight matrixs and $\mit{b_{i}} \in{\Bbb{R}^{C}}$, $\mit{n_{i}} \in{\Bbb{R}^{C/r}}$ are bias items. $\tau$  and $\mu$ mean softmax and ReLU activation function, respectively. Finally, the output result is the summary of softmax attention-weighted feature maps from two branches:
\begin{flalign}
 \tilde{x_{out}} = \sum_{i=1}^{2} \bf{F}_{scale}(\mit{s_{i}},\mit{x_{i})}=\sum_{i=1}^{2}  \mit{s_{i}} \otimes \mit{x_{i}},
\end{flalign}
 where $\bf{F}_{scale}(\mit{s_{i}}, \mit{x_{i}})$, refers to
channel-wise multiplication between the feature map and softmax weight in each branch.

The Conformer  \cite{conformer}  block is the combination of convolution and Transformer \cite{vaswani2017attention}, which has achieved state-of-art performance in ASR. It is composed of two feed-forward blocks that sandwich a multi-head self-attention (MHSA) block.  All these four blocks start with layer normalization operations. And the second feed-forward block is also followed by layer normalization. A residual connection is added behind each block. Assume $z$ is the input to Conformer, the output $o$ can be calculated through intermediate $\mathbf{z}^{\prime}$, $\mathbf{z}^{\prime}$ and $\mathbf{z}^{\prime \prime}$ as:
\begin{equation}
\hat{\mathbf{z}}=\mathbf{z}+\frac{1}{2} \mathrm{FFN}(\mathbf{z}),
\end{equation}
\begin{equation}
\mathbf{z}^{\prime}=\hat{\mathbf{z}}+\operatorname{MHSA}(\hat{\mathbf{z}}),
\end{equation}
\begin{equation}
\mathbf{z}^{\prime \prime}=\mathbf{z}^{\prime}+\operatorname{Conv}\left(\mathbf{z}^{\prime}\right),
\end{equation}
\begin{equation}
\mathbf{o}=\operatorname{Layernorm}\left(\mathbf{z}^{\prime \prime}+\frac{1}{2} \mathrm{FNN}\left(\mathbf{z}^{\prime \prime}\right)\right),
\end{equation}
where FFN(·), MHSA(·), Conv(·) and Layernorm(·) denote a feed-forward network block, a multi-head self-attention block, a convolution block, and a layer normalization layer, respectively.

Besides, multi-track ACCDOA is introduced into the proposed SELD model. The ACCDOA format is a SELD mechanism that converts DOA estimation into sound event class activity (as described in \ref{ssec:Multiaccdoa}), which has combined SED and DOA estimation into one branch and integrates two optimization branches into one regression optimization. By adding the DK module to adaptively extract features  and the conformer block to model local and global context dependencies, we propose a SELDnet-DK model that involves DK modules and the Conformer blocks. Additionally, we propose two SELD structures based on EINv2 framework \cite{9413473}, which has achieved state-of-art performance in polyphonic SELD tasks. The first system, shortened to EINv2-DK\#1, is modified as a structure for two input features and single-task optimization. The normal convolution blocks with soft parameter sharing blocks are retained,  then the outputs from two branches are concatenated and fed into the DK convolution modules and  the Conformer encoders. For the second one, shortened to EINv2-DK\#2, we preserve the original framework, applying DK modules and multi-track ACCDOA output.

\subsection{Multi-track ACCDOA}
\label{ssec:Multiaccdoa}
In complex spatial scenes, it's common that two sound sources occur in different positions while they belong to the same class. If simply using a frame-wise vector as the SED output format, the same-class polyphony with various DOA can not be presented clearly. Therefore, conventional SELD presentation degrades the SELD performance in real scenes with frequent polyphony. To overcome this challenge, we introduce a multi-track Activity-coupled Cartesian DOA(ACCDOA) vector \cite{Shimada2022} as the output format in our proposed system. ACCDOA uses the Cartesian information of the spatial sound event as the DOA presentation. Cartesian information of sound events is converted from the azimuth angle $\phi$ and elevation angle $\theta$ of the references via:
\begin{flalign}
&x = \cos{\phi}\cos{\theta}, \\
&y = \sin{\phi}\cos{\theta}, \\
&z = \sin{\theta}.
\end{flalign}
The sound event is active when $\Vert (x,y,z) \Vert$=1 where $\Vert\cdot\Vert$ is the L2 norm, The multi-track ACCDOA vector is formulated as :
\begin{flalign}
\textbf{P}_{nct} = \alpha_{nct}\textbf{R}_{nct},
\end{flalign}
where $\textbf{P}_{nct}$/$\textbf{R}_{nct}$ is the ACCDOA prediction/ground truth at track $n$, class $c$, and frame $t$. Once $\alpha_{nct}$ has reached the threshold, the corresponding class of sound event is activated. 

To overcome the overlap and permutation issues of polyphonic sound, multi-track ACCDOA with auxiliary duplicating
permutation invariant training(ADPIT) loss is introduced into the proposed system. Being different from track-wise PIT loss, an extra class-wise permutation is set for  the ACCDOA format, which can be written as follow:
\begin{equation}
\mathcal{L}^{\mathrm{PIT}}=\frac{1}{C T} \sum_c^C \sum_t^T \min _{\alpha \in \operatorname{Perm}(c t)} l_{\alpha, c t}^{\mathrm{ACCDOA}},
\end{equation}
\begin{equation}
l_{\alpha, c t}^{\mathrm{ACCDOA}}=\frac{1}{N} \sum_n^N \operatorname{MSE}\left(\boldsymbol{P}_{\alpha, n c t}^*, \hat{\boldsymbol{P}}_{n c t}\right),
\end{equation}
where ${\boldsymbol{P}_{\alpha, n c t}^*}$ is an ACCDOA target of a permutation $\alpha$, and ${\hat{\boldsymbol{P}}_{n c t}}$
is an ACCDOA prediction at track $n$, class $c$, and frame $t$. The loss for each track is computed using the MSE loss function.

The maximum of polyphonic sound is 3, i.e. T=3, When there are fewer active events than tracks in an event class, zero vectors are assigned as auxiliary targets of the inactive tracks in the class, which might interfere the optimization of multi-track ACCDOA training. To map the track with the same target, the inactive track is assigned with duplicated original active targets instead of zero vectors.
During inference, duplicated outputs are unified in two steps. First, a threshold is set to determine whether each output is active or not. Second, if there are two or more active sound events for the same class, the angle difference between them is calculated; if an angle difference between the outputs is smaller than a threshold, i.e., the outputs are similar, we take the average of the outputs to a unified output.

\subsection{Real scenes dedicated training strategy}
\label{ssec:strategy}
lt costs a lot of time and effort to make a SELD dataset of real spatial scenes. Being different from the synthetic dataset,  the strong SED labels (onset and offset) and DOA ground truths demand human annotation along the timeline. Therefore, the total duration of a real SELD dataset is significantly less than a synthetic SELD dataset. The reverberation, overlapped sound, moving sources, and other factors related to acoustic propagation determine the dynamic change of sound fields while the RIRs are static in the synthesis procedure. The spatial divergence between synthesis and real recordings has led to the fact: a SELD model can learn the pattern of inter-channel phase difference in direct-arrival waves and can be optimized using massive synthetic multichannel recordings, while it is not the best-optimized model for inference in real spatial scenes. 

To improve the performance of our SELD model in real scenes, we proposed two effective scene-dedicate training strategies:

\begin{itemize}
\item[-] Scene Transfer (ST). The model is first pre-trained and evaluated using only synthetic data and then trained with real data with a lower learning rate. It's a simple fine-tuning method from synthesis to reality. The model has acquired prior knowledge learned from synthetic data during training on the real training set.

\item[-] Scene Concentrate (SC). Features of two kinds of datasets are deemed homogeneous since they share the same absolute DOA information of direct wave. Therefore, they can be fed into models simultaneously and fine-tuned. 
Synthetic and real data are mixed in each mini-batch during model training. The first-step model is gathered when it achieves the best validation results on real test data. Next, the model is fine-tuned using only real spatial recordings with a lower learning rate.
\end{itemize}

\begin{figure}[h]
    \centering
     \subfigure[scene-transfer training]{\includegraphics[width=8.cm]{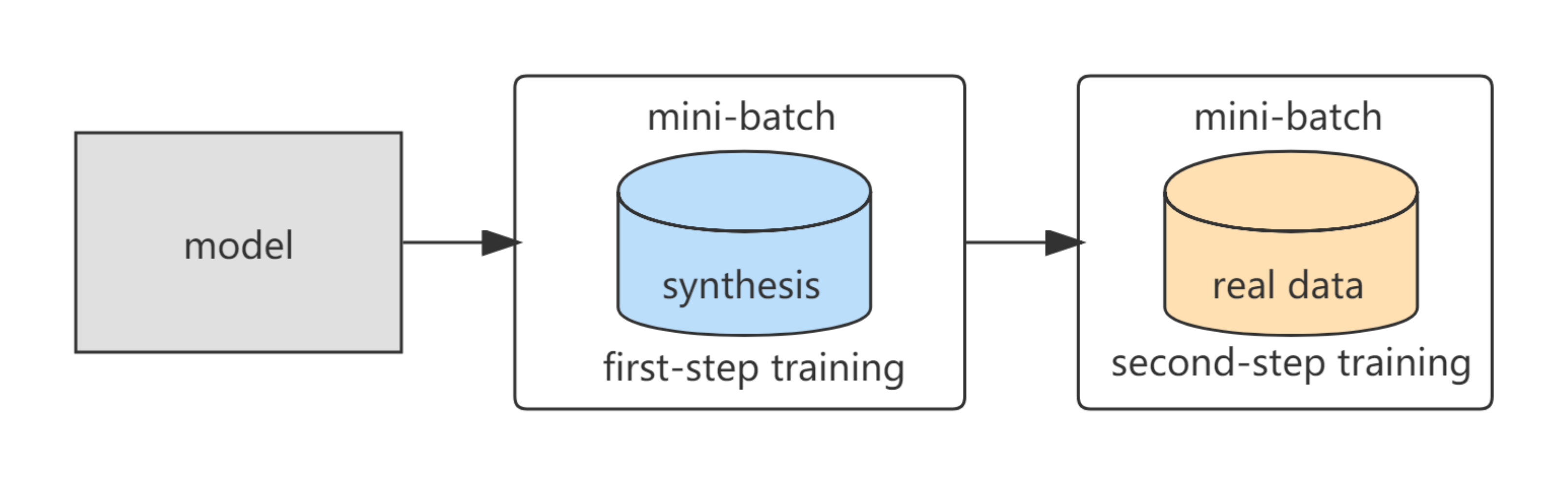}}
    \subfigure[scene-concentrate training]{\includegraphics[width=9.cm]{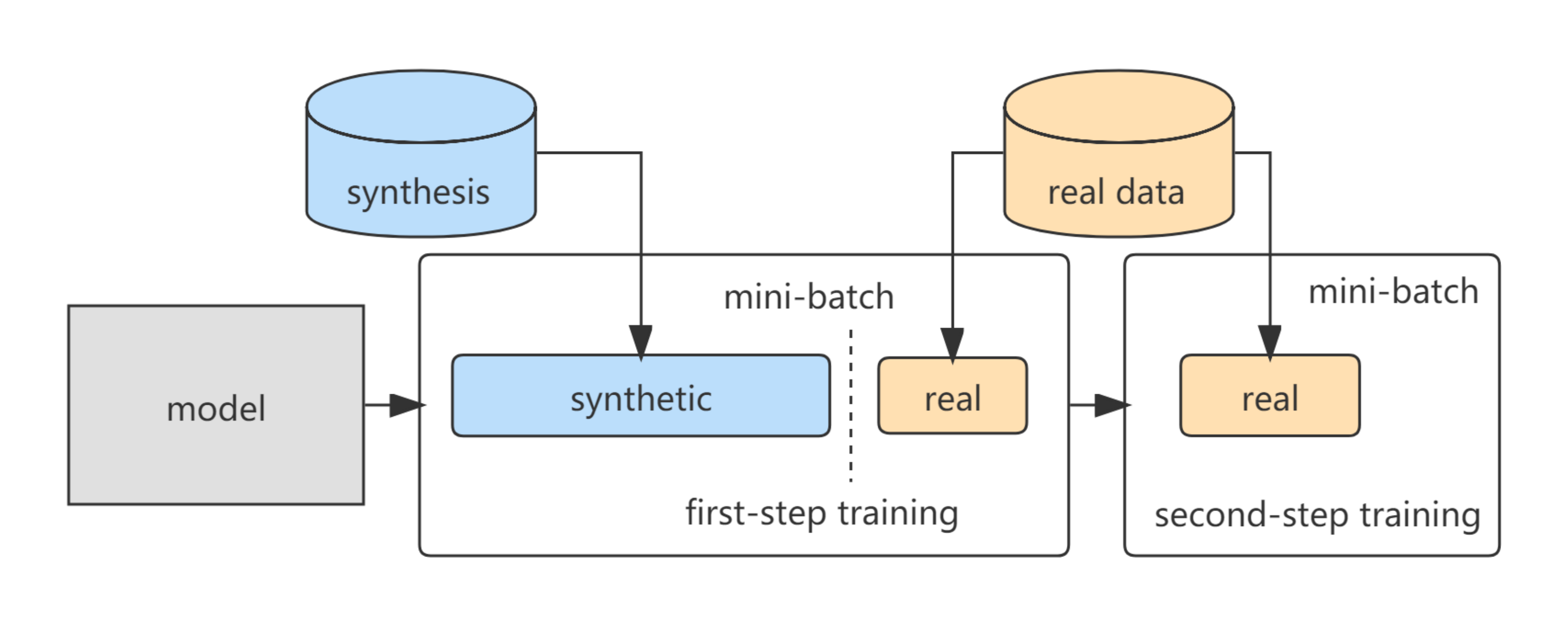}}
    \caption{Two training strategies.}
    \label{fig:workflow}
\end{figure}

\subsection{Data augmentation}
\label{ssec:DataAug}
In order to further increase the diversity of data and the generalization ability of the model, we introduce a data synthesis method to enlarge the variety of spatial recordings of the SELD dataset. This method synthesizes multi-channel recordings by convolving the monophonic recordings with spatial impulse responses, the sources of monophonic audio clips and room impulse responses along with noises are derived from FSD50K and TAU-SRIR DB. Moreover, we apply channel rotation \cite{fourstage} for data balancing. Specifically, we swapped channels of the multichannel data and change their DOA labels so that the  generated spatial recordings are from different directions. This method can maintain the original context of sound events while effectively creating new recordings with various spatial characteristics.
Tab. \ref{tab:mic_rotation}  describes the channel rotation methods used in our SELD systems. The workflow of these two methods is illustrated in Fig.  \ref{fig:workflow}.

\begin{figure}[t]
    \centering
  \subfigure[spatial data synthesis]{\includegraphics[width=4.2cm]{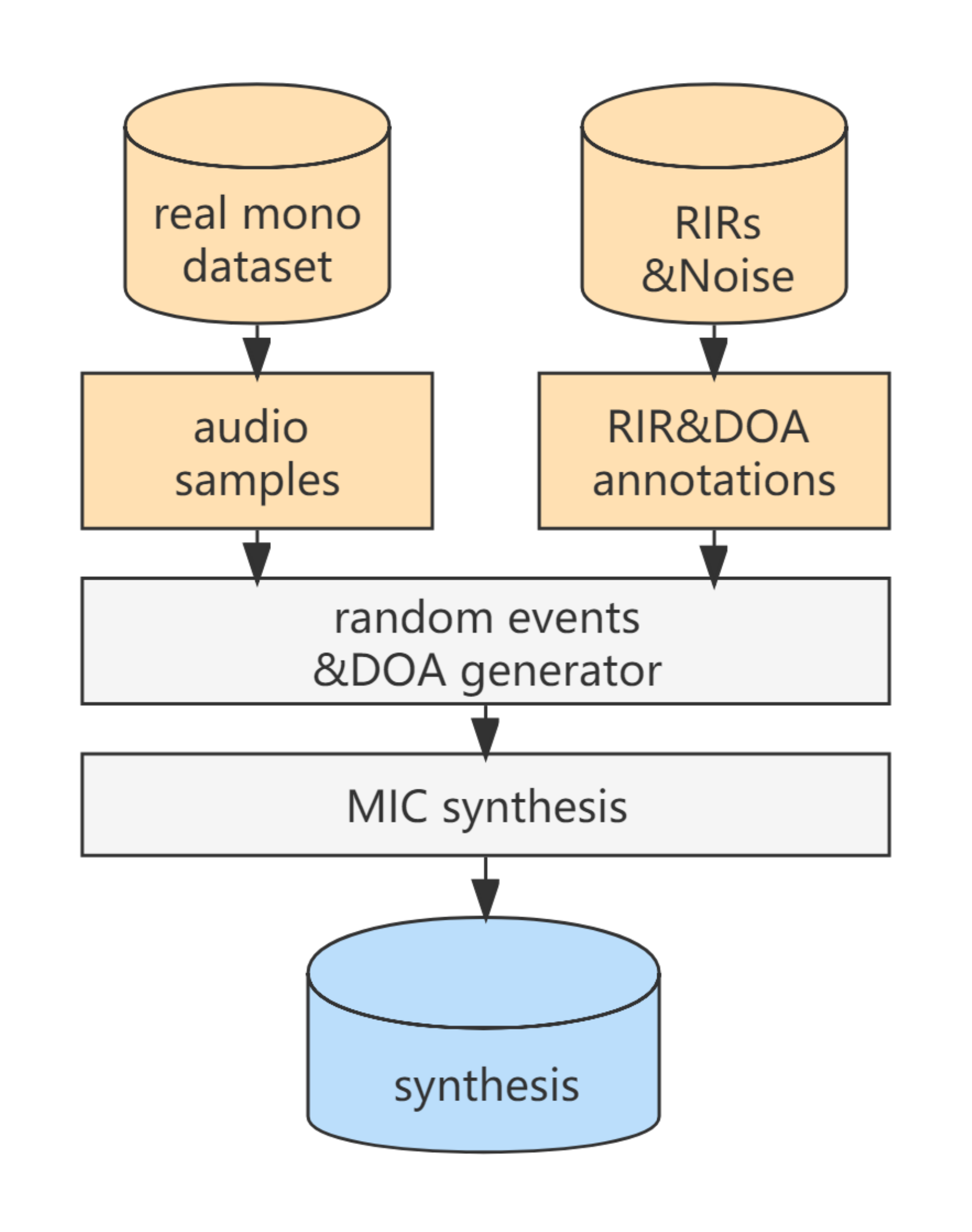}}
    \subfigure[channel rotation]{\includegraphics[width=4.cm]{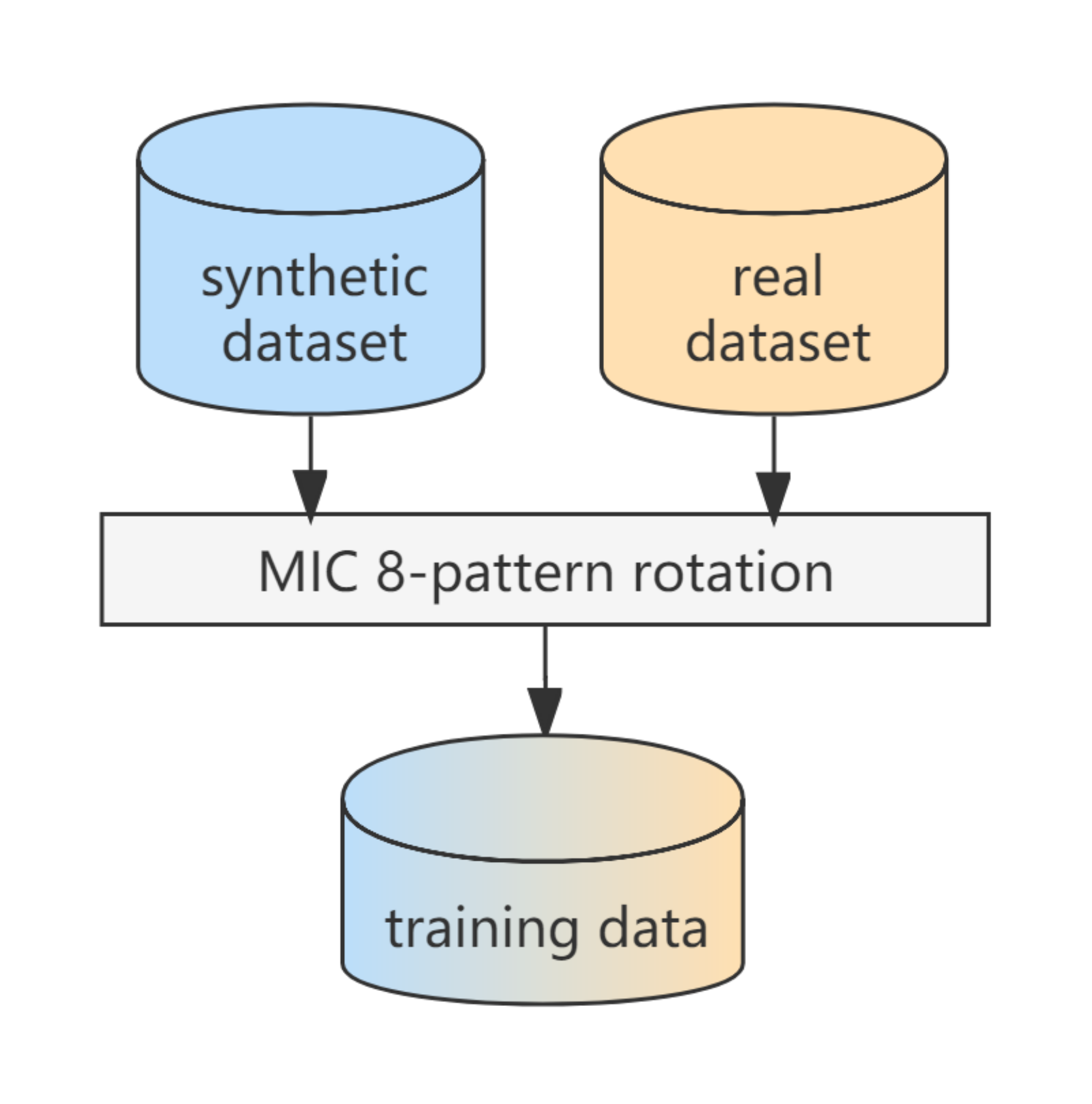}}
    \caption{Data extension workflow, spatial data synthesis(a) uses external audio data and spatial pulse responses and spatial noises, MIC channel rotation(b) is applied to existing spatial data.}
    \label{fig:workflow}
\end{figure}

We also introduce three data augmentation methods in our SELD system: FMix  \cite{Fmix}, mixup  \cite{mixup}, and Random cutout  \cite{cutout}. 
We use the mixing augmentation method, i.e., FMix and mixup, which have been widely used for environmental sound recognition. 
In the mixing augmentation method, the data and labels are mixed to generate new training data. 
Random cutout is used to mask areas of features without changing its original labels. In our experiment, we randomly mask the quarter of input features on each channel.

\begin{table}[t]
\centering
\caption{MIC data rotation/swapping original by \cite{fourstage}. $\phi$ and $\theta$ denote the azimuth angle and elevation angle, $C_{M}^{new}$ and $C_{M}$ denote the new channel and original channel.}
\label{tab:mic_rotation}
\renewcommand\arraystretch{1.25}
\setlength{\tabcolsep}{1mm}{
\begin{tabular}{cc}
\hline
DOA transformation & MIC channel swapping        \\ \hline
$\phi$ = $\phi$-$\pi$/2, $\theta$= -$\theta$  & $C_{1}^{new}$=$C_3$,$C_{2}^{new}$=$C_1$,$C_{3}^{new}$=$C_4$,$C_{4}^{new}$=$C_2$     \\ \hline
$\phi$ = -$\phi$-$\pi$/2, $\theta$=$\theta$    & $C_{1}^{new}$=$C_4$,$C_{2}^{new}$=$C_2$,$C_{3}^{new}$=$C_3$,$C_{4}^{new}$=$C_1$     \\ \hline
$\phi$ = $\phi$, $\theta$=$\theta$         & 
$C_{1}^{new}$=$C_1$,$C_{2}^{new}$=$C_2$,$C_{3}^{new}$=$C_3$,$C_{4}^{new}$=$C_4$ \\ \hline
$\phi$ = -$\phi$, $\theta$= -$\theta$      & 
$C_{1}^{new}$=$C_2$,$C_{2}^{new}$=$C_1$,$C_{3}^{new}$=$C_4$,$C_{4}^{new}$=$C_3$ \\ \hline
$\phi$ = $\phi$+$\pi$/2, $\theta$= -$\theta$   & 
$C_{1}^{new}$=$C_2$,$C_{2}^{new}$=$C_4$,$C_{3}^{new}$=$C_1$,$C_{4}^{new}$=$C_3$ \\ \hline
$\phi$ = -$\phi$+$\pi$/2, $\theta$= $\theta$   &
$C_{1}^{new}$=$C_1$,$C_{2}^{new}$=$C_3$,$C_{3}^{new}$=$C_2$,$C_{4}^{new}$=$C_4$ \\ \hline
$\phi$ = $\phi$+$\pi$, $\theta$= $\theta$     &
$C_{1}^{new}$=$C_4$,$C_{2}^{new}$=$C_3$,$C_{3}^{new}$=$C_2$,$C_{4}^{new}$=$C_1$ \\ \hline
$\phi$ = -$\phi$+$\pi$, $\theta$= -$\theta$    & 
$C_{1}^{new}$=$C_3$,$C_{2}^{new}$=$C_4$,$C_{3}^{new}$=$C_1$,$C_{4}^{new}$=$C_2$ \\ \hline
\end{tabular}
}
\end{table}

\subsection{Post-preocessing}
\label{ssec:ensemble}
Rotating-based inference, we generate inference data with different hop frames, then we sum the overlapping outputs with different weights. 
Furthermore, we rotate MIC data, estimate the multi-track ACCDOA vector, and rotate it back  \cite{postprocessing} with 8 rotation methods.
The final results are averaged on these outputs.

We averaged each track's outputs from different models trained using other augmentation methods and data. We set 0.4 as the threshold of the activity of sound events, when the length of the multi-track ACCDOA vector reaches this threshold, its corresponding sound event is deemed active.

\begin{table*}[]
\centering
\begin{threeparttable}[]

\caption{The submitted SELD systems to DCASE 2022 task3. The detailed configurations of each submitted system are listed according to their submitted technical reports.}
\label{tab:overview}
\begin{tabular}{l|llll|l}
\hline
 & Wu\_NKU's \cite{Wu_NKU_task3_report} & Official baseline & Wang\_SJTU's \cite{Wang_SJTU_task3_report} & Kang\_KT's \cite{Kang_KT_task3_report} & our systems \\ \hline
Structures & E-CRNN & SELDnet & \begin{tabular}[c]{@{}l@{}}CNN-MHSA\\ CNN-Transformer\\ w/{\&}w/o BiGRU\end{tabular} & \begin{tabular}[c]{@{}l@{}}CRNN multi-task\\ CRNN multi-ACCDOA \\ EINv2 multi-task\end{tabular} & \begin{tabular}[c]{@{}l@{}}SELDnet-DK\\ EINv2-DK\#1\\ EINv2-DK\#2\end{tabular} \\
Params & 10M & 606K & 672M & 64M & 64.07M \\
Ensemble & \ding{55} & \ding{55} & \ding{52} & \ding{52} & \ding{52} \\
Feature & \begin{tabular}[c]{@{}l@{}}log-Mel spec\\GCC-PHAT+VQT\end{tabular} & \begin{tabular}[c]{@{}l@{}}log-Mel spec\\ GCC-PHAT\end{tabular} & \begin{tabular}[c]{@{}l@{}}log-Mel spec \\ GCC-PHAT\end{tabular} &\begin{tabular}[c]{@{}l@{}} SALSA-Lite \\ log-Mel spec + IVs \end{tabular}& SALSA-Mel \\
channel×bins & 10×64 & 10×64 & 10x64 & 7×384/7×64 & 7×64 \\ \hline
Extra tricks & - & - & domain adaption & - & scenes dedicated \\
Optimizer & Adam & Adam & - & AdamW & Adam \\
MaxLR & 1.0E-3 & 1.0E-3 & - & 1.0E-3 & 3.0E-4 \\ \hline
External dataset & FSD50K & FSD50K & AudioSet & AudioSet & FSD50K \\
Pre-trained model & \ding{55} & \ding{55} & PANN & \ding{55} & \ding{55} \\
Synthetic mixture & $\sim20h$ & $\sim20h$ & $\ge20h^*$ & $\ge20h^*$ & $\sim50h$ \\
Rotation/swapping & \ding{55} & \ding{55} & \ding{55} & \ding{52} & \ding{52} \\
masking & \ding{55} & \ding{55} & \ding{55} & SpecAugment + cutout & cutout \\
Mixing & mixup & \ding{55} & \ding{55} & \ding{55} & Fmix+mixup \\
Frequency shift & \ding{55} & \ding{55} & \ding{55} & \ding{52} & \ding{55} \\ \hline
{$ER_{20}$°↓} & 0.69 & 0.61 & 0.67 & \textbf{0.47} & \textbf{0.47} \\
{$F_{20}$°↑} & 17.9 & 21.6 & 27.0 & 45.9 & \textbf{49.3} \\
{$LE_{CD}$↓} & 28.5° & 25.9° & 24.4° & \textbf{15.8°} & 16.9° \\
{$LR_{CD}$↑} & 44.5 & 48.1 & 60.3 & 59.3 & \textbf{67.9} \\
{$\varepsilon_{SELD}$↓} & 0.538 & 0.514 & 0.483 & 0.376 & \textbf{0.348} \\ \hline
\end{tabular}
 \begin{tablenotes}
       \item [-] None or not mentioned in the reports.
       \item [*] The source description is ambiguous.
     \end{tablenotes}
\end{threeparttable}
\end{table*}

\section{EXPERIMENTS}
\label{sec:exp}

In this section, wepresent the SELD experiments on the public dataset of DCASE 2022 Task3. First of all,  we introduce the dataset used in our experiments including FSD50K, STARSS22 and TAU-SRIR DB. Secondly, we introduce the baseline and state-of-the-art systems in DCASE SELD task. Finally, we present the experimental results and discussions of the SELD systems.

\subsection{Dataset}
\label{ssec:dataset}
We train our models using the synthesis original from  FSD50K and the real data from the development set of STARSS22. Models are evaluated using the evaluation set of STARSS22.

\text {FSD50K dataset} \cite{fonseca2022FSD50K}: This dataset is an open dataset of human-labeled sound events, which contains over 51k audio clips totaling over 100h of audio manually labeled using 200 classes drawn from the AudioSet Ontology. The audio clips are gathered from Freesound and are licensed under Creative Commons licenses, which allow easy sharing and reuse.

\textbf{STARSS22 dataset} \cite{star}: This dataset is collected in two different countries, in Tampere, Finland by the Audio Research Group of Tampere University (TAU), and in Tokyo, Japan by SONY. 13 target classes are identified in the recordings and strongly annotated by humans. Spatial annotations for those active events are captured by an optical tracking system. This dataset contains multichannel recordings of sound scenes in various rooms and environments, totaling 5h in the development set and 2h in the blind test set. The blind test is used to evaluate the submitted SELD systems in DCASE 2022 task 3.

\textbf{TAU-SRIR DB dataset} \cite{politis2020dataset}: This dataset contains spatial room impulse responses (SRIRs) captured in various spaces of TAU, Finland, for a fixed receiver position and multiple source positions per room, along with separate recordings of spatial ambient noise captured at the same recording point. The dataset can be used to generate multichannel mixtures for localization, multichannel speech enhancement, and acoustic scene analysis. Monophonic audio files and different SRIRs are used to generate synthetic spatial audio files.

\textbf{2022 DCASE SELD task released SELD synthsis} \cite{politis2021dataset}: This dataset consists of synthetic spatial audio mixtures of sound events spatialized for two different spatial formats using real measured room impulse responses measured in various spaces of Tampere University (TAU). It contains 20h synthetic data including 13 target sound classes and their presence is similar to targets in real recordings. These synthetic recordings 
 is released as public data support for training a baseline SELD system.

In our proposed method, we use FSD50K as the source of synthetic data. Therefore, multiple datasets are used and combined in our experiments.

\subsection{Baseline system}
\label{ssec:baseline}

\textbf{Official baseline \cite{star}} It is a straightforward convolutional recurrent neural network (CRNN), shortened to SELDnet. In the 2022 DCASE SELD task, a few modifications have been integrated into the model. A Multi-head self-attention block is applied after the bidirectional GRU block.  The whole system is trained using a mixture of released synthetic and realistic multichannel recordings,  split from synthetic recordings generated from FSD50K (fold 1,2, $\sim$20h) and the development set of STARSS22 (fold 3, $\sim$2.5h), respectively. 

\textbf{Wu\_NKU's \cite{Wu_NKU_task3_report}:} It is a system based on an MLP-Mixer enhanced convolutional recurrent neural networks with improved input features with variable-Q transform.

\textbf{Wang\_SJTU's \cite{Wang_SJTU_task3_report}:} It is a model combination of  CRNN, CRNN with multi-head self-attention (MHSA) technique, and CNN-Transformer encoder. Additionally, to overcome the domain discrepancy led from real sound space, they present their SELD works with the large pre-trained model (PANN) active learning and domain adaption.

\textbf{Kang\_KT's \cite{Kang_KT_task3_report}:} They provide a track-wise ensemble of CRNNs with different output formats. Their models are trained using two data formats, including first-order ambisonic data and microphone array data.

\subsection{Experimental setups}
\label{ssec:subexp}

The SELD systems are evaluated on the evaluation set of STARSS22.
Five metrics are used for evaluation \cite{politis2020overview}: error rate ($ER_{20}$°), F-score ($F_{20}$°), $LE_{CD}$, $LR_{CD}$, $\varepsilon_{SELD}$, where $\varepsilon_{SELD}$ is an aggregated metric of the SELD system's overall performance, and computed as :
\begin{align}
\varepsilon_{SELD}\!=\!\frac{1}{4}[ER_{20\degree}\!+\!(\!1\!-\!F_{20\degree}\!)\!+\!\frac{LE_{CD}}{180\degree}\!+\!(1-LR_{CD})]
\end{align}

During feature extraction, the sampling frequency is set to 24kHz, the number of Mel filters is 64, and the frame length and hop length of STFT is 40ms and 20ms respectively.. The length of the input is 250 frames (5s). 

For the models, the kernel size of each convolution block is set as  [64, 128, 256] and the maximum of 256. We add two Conformer blocks with a hidden size of 256 that fit the size of the feature map from the CNN.

For the scenes-dedicated training, the model is firstly trained on synthetic data for 100 epochs with a learning rate of 0.0005 using the Adam optimizer \cite{kingma2014adam}. 
The model with the best SELD score is saved and further trained on real recordings for extra 25 epochs with a learning rate of 0.1 decay.We use a batch size of 64.

By applying data augmentation, we generate new synthesis data, and the amount of recordings is extended to about 20h using channel rotation. Furthermore,  training augmentation methods  such as Fmix (FM), mixup (MIX), and random cutout (RC) are applied throughout the whole process of training. All experiments are performed using PyTorch \cite{paszke2017automatic} toolkit, on a single GeForce RTX2080Ti GPU.

\subsection{Results and discussions}
\label{ssec:subexp}

\subsubsection{The overall result}

Table \ref{tab:overview} shows the details and SELD systems evaluated on the blind test set of DCASE 2022 Task3. As  shown in the table, our proposed method outperforms the official baseline by a large margin. The proposed systems can reach a considerable performance of $\varepsilon_{SELD}$. Training strategies and augmentations also play an important role in SELD systems.

\begin{table}[h]
\centering
\caption{The comparison of the three proposed structures.  All systems are trained with SC training strategy.}
\label{tab:structure}
\resizebox{\columnwidth}{!}{%
\begin{tabular}{llllll}
\hline
System & {$ER_{20}$°↓} & {$F_{20}$°↑} & {$LE_{CD}$↓} & {$LR_{CD}$↑} & {$\varepsilon_{SELD}$↓} \\ \hline
\multicolumn{1}{l|}{SELDnet(baseline) + Data Aug.} & 0.65 & 26.1 & 22.4° & 49.0 & 0.507 \\
\multicolumn{1}{l|}{+ Conformer encoder rep. BiGRU} & 0.57 & 40.8 & 21.9° & 61.2 & 0.412 \\
\multicolumn{1}{l|}{+ DK module} & 0.50 & 49.3 & 18.2° & 57.6 & 0.383 \\
\multicolumn{1}{l|}{+ EINv2-DK\#1} & 0.50 & 45.4 & 18.1° & 60.3 & 0.385 \\
\multicolumn{1}{l|}{+ EINv2-DK\#2} & 0.50 & 45.4 & 17.5° & 59.0 & 0.388 \\
\multicolumn{1}{l|}{+ ensemble of proposed systems} & \textbf{0.44} & \textbf{54.2} & \textbf{16.0°} & \textbf{65.4} & \textbf{0.333} \\ \hline
\end{tabular}%
}
\end{table}

\subsubsection{Results of modules}
\label{ssec:model exp}
Table \ref{tab:structure} shows the SELD performances with respect to  different structures introduced in our method. First, we replace GRU and self-attention modules with two Conformer blocks. Next, we incorporate DK module into downsampling CNN. Moreover, we implement soft parameter-sharing units in two-branch networks.
As a result, CNN-Conformer has achieved an  overwhelmingly better SELD performance thanks to the Conformer encoder that models local and global features. Based on the CNN-conformer-like structure, SELDnet-DK incorporates the DK module into the last  convolution block. The selective kernel has shown its event-aware modeling ability so that SELDnet-DK is the most sensitive and achieves the highest $F_{20}$° among single models. In two EINv2-based structures, soft-parameter sharing blocks fuse time-frequency features  and spatial features in multi-scales. It works effectively in multi-branch prediction, resulting in a similar SELD score as SELDnet-DK, tending to lower $LE_{CD}$ and higher $LR_{CD}$. Overall,  each module involved contributes to the SELD task and accumulates improvements by giving a  series of integration. Finally, the ensemble of three proposed systems achieves the best SELD score on the development set.

\begin{table}[h]
\centering
\caption{The performance of different training strategies. - and * denote training only using real and synthetic data respectively. ST and SC denote Scene Transfer, Scene Concentrated, respectively }
\label{tab:strategy}
\resizebox{\columnwidth}{!}{%
\begin{tabular}{ll|lllll}
\hline
System & TS & {$ER_{20}$°↓} & {$F_{20}$°↑} & {$LE_{CD}$↓} & {$LR_{CD}$↑} & {$\varepsilon_{SELD}$↓} \\ \hline
\multirow{4}{*}{SELDnet} & - & 0.72 & 11.0 & 109.1° & 18.0& 0.760 \\& * & 0.69 & 22.0 & 29.7° & \textbf{43.0} & 0.560 \\
 & ST & 0.67 & 21.3 & 25.8° & 34.6 & 0.564 \\
 & SC & \textbf{0.65} & \textbf{26.6} & \textbf{21.9}° & 39.6 & \textbf{0.536} \\ \hline
\multirow{4}{*}{SELDnet-DK} & - & 0.77 & 15.6 & 68.9° & 21.2& 0.698 \\& * & 0.55 & 43.8 & 19.8° & 55.2 & 0.422 \\
 & ST & 0.53 & 41.8 & 19.4° & 57.2 & 0.413 \\
 & SC & \textbf{0.50} & \textbf{49.3} & \textbf{18.2°} & \textbf{57.6} & \textbf{0.383} \\ \hline
\end{tabular}%
}
\end{table}

\subsubsection{Results of scene-dedicate training}
\label{ssec:strategy exp}
 We evaluate the proposed scene-dedicate training strategies on two different architectures.  As shown in Tabel \ref{tab:strategy}, scene concentrate (SC) achieves a significant improvment compared with the other two training methods. we assume that the model is supposed to be best optimized to the interested spatial scenes. The conclusion might be extended to the acoustic-related scene-to-scene, class-to-class scenario, and other domain generation in future work.

\begin{table}[h]
\centering
\caption{Preliminary experiments using different features without augmentation}
\label{tab:pre-exp}
\resizebox{\columnwidth}{!}{%
\begin{tabular}{lllllll}

\hline
Feature & \multicolumn{1}{l|}{Input size}    &{$ER_{20}$°↓} & {$F_{20}$°↑ } & {$LE_{CD}$↓} & {$LR_{CD}$↑ } & {$\varepsilon_{SELD}$↓} \\ \hline
log-Mel + GCC              & \multicolumn{1}{c|}{[7,128]} & 0.71                 & 18                   & 32.2°                 & \textbf{47.0}           &0.560                        \\
 SALSA-Lite              & \multicolumn{1}{c|}{[7,382]} & 0.69                 & 21.8                   & \textbf{29.3°}                 & 43.7           &\textbf{0.550}                        \\
SALSA-Mel              & \multicolumn{1}{c|}{[7,64]}  & \textbf{0.69}                 & \textbf{22.0}                   & 29.7°                 & 43.0       &0.560                          \\ \hline
\multicolumn{1}{l}{} & \multicolumn{1}{l}{}            & \multicolumn{1}{l}{} & \multicolumn{1}{l}{} & \multicolumn{1}{l}{} & \multicolumn{1}{l}{} & \multicolumn{1}{l}{}
\end{tabular}%
}
\end{table}

\subsubsection{Results of input feature}
\label{ssec:feature exp}
To verify the effectiveness of the proposed feature, we train the baseline system using different features. As shown in Tabel \ref{tab:pre-exp}, compared with log-Mel+GCC feature, SALSA-Lite gains lower  $ER_{20}$°,$LE_{CD}$, higher $F_{20}$°, and SELD performance.
The proposed feature, i.e., SALSA-Mel, achieves 0.01 degradation on the SELD performance compared to SALSA-Lite, while it is one-quarter the input size of the original SALSA-Lite, which makes it suitable for developing a low-complexity model in low-resource situations.

\begin{table}[h]
\caption{Comparison of different augmentation methods. We are using SELDnet and  SALSA-Mel as our baseline. In this table, the generation means adding synthetic data and rotation means 8 MIC channel rotation. ALL means training with all augmentation methods.}
\label{tab:augment}
\resizebox{\columnwidth}{!}{%
\begin{tabular}{ll|lllll}
\hline
System &Data Aug.& {$ER_{20}$°↓} & {$F_{20}$°↑ } & {$LE_{CD}$↓} & {$LR_{CD}$↑ }  & {$\varepsilon_{SELD}$↓} \\ \hline
\multirow{8}{*}{SELDnet} &w/o& 0.69 & 20.0 & 29.4° & 43.0 & 0.560 \\ 
   &synthesis(S) & 0.69 & 25.5 & 26.3° & 44.6 &0.534 \\ 
  &rotation(R) & 0.65 & 25.1 & 23.6° & 44.6 &0.521 \\
   &FMix(FM) & 0.66 & 21.7 & 25.6° & 37.2 &0.553\\ 
   &mixup(MIX) & 0.72 & 20.7 & 28.8° & 44.6 &0.559\\
  &SpecAug(SA) & 0.68 & 23.0 & 36.5° & 36.5 &0.572\\
  &RandomCutout(RC) & 0.65 & 22.7 & 29.4 & 42.1&0.541 \\ 

\multicolumn{1}{c}{}   &ALL w/o SA & \textbf{0.65} & \textbf{26.1} & \textbf{22.4°} & \textbf{49.0} &\textbf{0.507} \\ \hline
\multirow{3}{*}{SELDnet-DK}&S& 0.58 & 39.1 & 19.9° & 62.0 & 0.421 \\
&S+RC& 0.57 & 43.3 & 20.7° & \textbf{65.8} & \textbf{0.399} \\
&S+RC+MIX& \textbf{0.54} & \textbf{44.0} & \textbf{19.2°} & 59.8 & 0.404 \\ \hline

\multirow{3}{*}{EINv2-DK\#1}&S & 0.58 & 34.1 & 20.6° & 59.6 & 0.440 \\
&S+RC & 0.54 & 41.6 & 19.9° & \textbf{61.8} & \textbf{0.405} \\
&S+RC+MIX & \textbf{0.53} & \textbf{41.8} & \textbf{17.7°} & 53.9 & 0.419 \\ \hline

\multirow{3}{*}{EINv2-DK\#2}&S & 0.60 & 38.6 & 21.6° & 60.2 & 0.434 \\
&S+RC & \textbf{0.55} & \textbf{43.4} & \textbf{18.7°} & \textbf{62.9} & \textbf{0.399} \\
&S+RC+MIX & 0.59 & 37.0 & 20.6° & 51.6 & 0.455 \\ \hline
\end{tabular}%
}
\end{table}

\subsubsection{Results of augmentation}
\label{ssec:aug exp}
Tabel \ref{tab:augment} shows the SELD performance of using different augmentation methods on the SELDnet, SELDnet-DK and EINV2-DKs. First, we do ablation experiments for study the contribution of different data augmentation methods on SELDnet. Synthesis achieves the highest F-score and we assume that the dataset is more balanced by improving the presence of the scarce classes As shown in this table, channel rotation achieves the lowest $LE_{CD}$, which is the most effective spatial augmentation method. Mixing methods such as FMix and mixup slightly improve the performance of SELD tasks. Random cutout also improves $F_{20}$°, $LE_{CD}$ effectively. we also experiment with an effective masking method SpecAugment which has been commonly used in ESR methods \cite{park2019specaugment}. SpecAugment improves the SED results but degrades the localization results, leading to worse SELD performance in general. Similar experimental results were also found by  Shimada et al \cite{spatialmixup}. We assume that SpecAugment masks available temporal information to handle a frame-wise output task. Therefore, we do not take SpecAugment in our SELD systems. The baseline model incorporates all data augmentation methods except for SpecAugment and finally achieves a SELD score of 0.507.

The performances of the proposed systems with different augmentation methods are listed in Table \ref{tab:augment}. In this table, our systems with extension (synthesis and channel rotation) outperform the baseline model with all augmentation methods. What is more, it shows various improvements with Random cutout and mixing (FMix and mixup). For both three proposed models, Random cut increases $LR_{CD}$ , and mixing methods decrease {$ER_{20}$°}, {$LE_{CD}$}, while sacrificing $LR_{CD}$. We conclude the effectiveness that three types of augmentation methods bring to our system: synthesizing new data, channel rotation i.e extension increases the general amount of physically legal data, mixing including FMix, mixup that generates virtual samples, covers more situations 
to make the SELD system robust, masking as random cutout masks part of the information of the feature to increase the Recall.

\begin{figure}[t]
    \includegraphics[width=9cm]{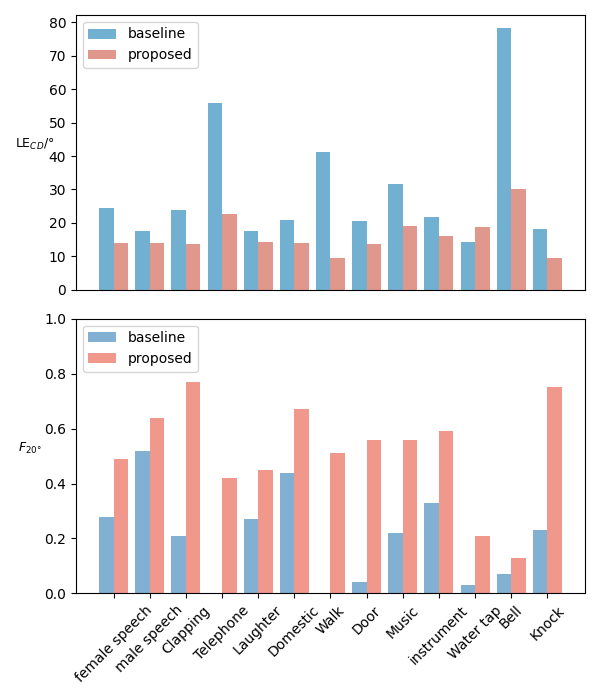}
    \caption{The comparison of the class-wise performance between the official baseline and the proposed system.}
    \label{fig:class}
\end{figure}

\subsubsection{Class-wise result}
The class-wise results are illustrated in Fig \ref{fig:class}. We can observe the incredible improvements in Telephone, Walk, Door, Water tap, and Bell. In the baseline system, the single digits in $F_{20}$° , high {$LE_{CD}$} for Telephone and Bell become cumbersome for the general metrics. 
However, in our system, $F_{20}$° and $LR_{CD}$ for each class has increased effectively, {$LE_{CD}$} for Bell has decreased from 78.31° to 30.25°, for Walk has decreased from 41.34° to 9.54°, accompanied with higher {$F_{20}$°} . ALL these improvements in each class bring better performance in general.

\section{Conclusion}
\label{sec:conclusion}
In this paper, we propose a Dynamic Kernel Convolution Network-based SELD system with Scene-dedicate Training strategy, which shows effective performance in real spatial sound scenes. In our system, we use log-Mel spectrogram and NIPD as the input feature.
We apply DK convolution modules and Conformer block into SELD models, then incorporate different networks with DK convolution modules based on multi-track ACCDOA systems and EINv2 systems. Moreover,
we synthesize multichannel audio files with external audio data and spatial impulse response data. A set of data augmentation methods are also used during network training.
Considering the difference between synthetic spatial audio and real recordings in exclusive environments, we propose two scene-dedicate training strategies. 
Finally, to further enhance the generalization ability of our system, we perform model ensemble to further enhance
the generalization ability of our system for SELD. The evaluations on the
dataset of DCASE2022 Task3 illustrated that our proposed system significantly outperforms the baseline system. 
For the submission to the 2022 DCASE SELD task, our best official system has achieved {$ER_{20}$} of 0.47, {$F_{20}$°} of 45.9, {$LE_{CD}$} of 16.9°, {$LR_{CD}$} of 67.9, which shows a robust and overwhelming performance for SELD system in real spatial environment.


%



\section*{Acknowledgment}

The authors would like to thank Tianlong Kong et al.'s novel idea and the application of the dynamic multi-scale convolution network.

\ifCLASSOPTIONcaptionsoff
  \newpage
\fi



\bibliographystyle{IEEEtran}
 \bibliography{refs}

\begin{thebibliography}{10}
\providecommand{\url}[1]{#1}
\def\UrlFont{\rmfamily}
\providecommand{\newblock}{\relax}
\providecommand{\bibinfo}[2]{#2}
\providecommand\BIBentrySTDinterwordspacing{\spaceskip=0pt\relax}
\providecommand\BIBentryALTinterwordstretchfactor{4}
\providecommand\BIBentryALTinterwordspacing{\spaceskip=\fontdimen2\font plus
\BIBentryALTinterwordstretchfactor\fontdimen3\font minus
  \fontdimen4\font\relax}
\providecommand\BIBforeignlanguage[2]{{%
\expandafter\ifx\csname l@#1\endcsname\relax
\typeout{** WARNING: IEEEtran.bst: No hyphenation pattern has been}%
\typeout{** loaded for the language `#1'. Using the pattern for}%
\typeout{** the default language instead.}%
\else
\language=\csname l@#1\endcsname
\fi
#2}}

\bibitem{Surveillance}
``\BIBforeignlanguage{English}{Audio surveillance of roads: A system for
  detecting anomalous sounds},'' \emph{\BIBforeignlanguage{English}{Ieee
  transactions on intelligent transportation systems}}, vol.~17, no.~1, pp.
  279--288, Jan. 2016.

\bibitem{stowell2016bird}
D.~Stowell, M.~Wood, Y.~Stylianou, and H.~Glotin, ``Bird detection in audio: a
  survey and a challenge,'' in \emph{2016 IEEE 26th International Workshop on
  Machine Learning for Signal Processing (MLSP)}.\hskip 1em plus 0.5em minus
  0.4em\relax IEEE, 2016, pp. 1--6.

\bibitem{blauert1997spatial}
J.~Blauert, \emph{Spatial hearing: the psychophysics of human sound
  localization}.\hskip 1em plus 0.5em minus 0.4em\relax MIT press, 1997.

\bibitem{Adavanne2018_JSTSP}
\BIBentryALTinterwordspacing
S.~Adavanne, A.~Politis, J.~Nikunen, and T.~Virtanen, ``Sound event
  localization and detection of overlapping sources using convolutional
  recurrent neural networks,'' \emph{IEEE Journal of Selected Topics in Signal
  Processing}, vol.~13, no.~1, pp. 34--48, March 2018. [Online]. Available:
  \url{https://ieeexplore.ieee.org/abstract/document/8567942}
\BIBentrySTDinterwordspacing

\bibitem{l3das}
E.~Guizzo, C.~Marinoni, M.~Pennese, X.~Ren, X.~Zheng, C.~Zhang, B.~Masiero,
  A.~Uncini, and D.~Comminiello, ``L3das22 challenge: Learning 3d audio sources
  in a real office environment,'' in \emph{ICASSP 2022 - 2022 IEEE
  International Conference on Acoustics, Speech and Signal Processing
  (ICASSP)}, 2022, pp. 9186--9190.

\bibitem{fonseca2022FSD50K}
E.~Fonseca, X.~Favory, J.~Pons, F.~Font, and X.~Serra, ``{FSD50K}: an open
  dataset of human-labeled sound events,'' \emph{IEEE/ACM Transactions on
  Audio, Speech, and Language Processing}, vol.~30, pp. 829--852, 2022.

\bibitem{politis2020dataset}
A.~Politis, S.~Adavanne, and T.~Virtanen, ``A dataset of reverberant spatial
  sound scenes with moving sources for sound event localization and
  detection,'' \emph{arXiv preprint arXiv:2006.01919}, 2020.

\bibitem{star}
A.~Politis, K.~Shimada, P.~Sudarsanam, S.~Adavanne, D.~Krause, Y.~Koyama,
  N.~Takahashi, S.~Takahashi, Y.~Mitsufuji, and T.~Virtanen, ``Starss22: A
  dataset of spatial recordings of real scenes with spatiotemporal annotations
  of sound events,'' \emph{arXiv preprint arXiv:2206.01948}, 2022.

\bibitem{DK}
T.~Kong, S.~Yin, D.~Zhang, W.~Geng, X.~Wang, D.~Song, J.~Huang, H.~Shi, and
  X.~Wang, ``{Dynamic Multi-Scale Convolution for Dialect Identification},'' in
  \emph{Proc. Interspeech 2021}, 2021, pp. 3261--3265.

\bibitem{SK}
X.~Zheng, Y.~Song, I.~McLoughlin, L.~Liu, and L.-R. Dai, ``An improved mean
  teacher based method for large scale weakly labeled semi-supervised sound
  event detection,'' in \emph{ICASSP 2021 - 2021 IEEE International Conference
  on Acoustics, Speech and Signal Processing (ICASSP)}, 2021, pp. 356--360.

\bibitem{cao2019polyphonic}
Y.~Cao, Q.~Kong, T.~Iqbal, F.~An, W.~Wang, and M.~D. Plumbley, ``Polyphonic
  sound event detection and localization using a two-stage strategy,'' in
  \emph{Acoustic Scenes and Events 2019 Workshop (DCASE2019)}, 2019, p.~30.

\bibitem{seld_TDNNF}
Q.~Wang, H.~Wu, Z.~Jing, F.~Ma, Y.~Fang, Y.~Wang, T.~Chen, J.~Pan, J.~Du, and
  C.-H. Lee, ``A model ensemble approach for sound event localization and
  detection,'' in \emph{2021 12th International Symposium on Chinese Spoken
  Language Processing (ISCSLP)}, 2021, pp. 1--5.

\bibitem{he2016deep}
K.~He, X.~Zhang, S.~Ren, and J.~Sun, ``Deep residual learning for image
  recognition,'' in \emph{Proceedings of the IEEE conference on computer vision
  and pattern recognition}, 2016, pp. 770--778.

\bibitem{chollet2017xception}
F.~Chollet, ``Xception: Deep learning with depthwise separable convolutions,''
  in \emph{Proceedings of the IEEE conference on computer vision and pattern
  recognition}, 2017, pp. 1251--1258.

\bibitem{Povey2018SemiOrthogonalLM}
D.~Povey, G.~Cheng, Y.~Wang, K.~Li, H.~Xu, M.~A. Yarmohammadi, and
  S.~Khudanpur, ``Semi-orthogonal low-rank matrix factorization for deep neural
  networks,'' in \emph{Interspeech}, 2018.

\bibitem{cao2020event}
Y.~Cao, T.~Iqbal, Q.~Kong, Y.~Zhong, W.~Wang, and M.~D. Plumbley,
  ``Event-independent network for polyphonic sound event localization and
  detection,'' \emph{arXiv preprint arXiv:2010.00140}, 2020.

\bibitem{accdoa}
K.~Shimada, Y.~Koyama, N.~Takahashi, S.~Takahashi, and Y.~Mitsufuji, ``Accdoa:
  Activity-coupled cartesian direction of arrival representation for sound
  event localization and detection,'' in \emph{ICASSP 2021 - 2021 IEEE
  International Conference on Acoustics, Speech and Signal Processing
  (ICASSP)}, 2021, pp. 915--919.

\bibitem{9413473}
Y.~Cao, T.~Iqbal, Q.~Kong, F.~An, W.~Wang, and M.~D. Plumbley, ``An improved
  event-independent network for polyphonic sound event localization and
  detection,'' in \emph{ICASSP 2021 - 2021 IEEE International Conference on
  Acoustics, Speech and Signal Processing (ICASSP)}, 2021, pp. 885--889.

\bibitem{vaswani2017attention}
A.~Vaswani, N.~Shazeer, N.~Parmar, J.~Uszkoreit, L.~Jones, A.~N. Gomez,
  {\L}.~Kaiser, and I.~Polosukhin, ``Attention is all you need,''
  \emph{Advances in neural information processing systems}, vol.~30, 2017.

\bibitem{app12105075}
\BIBentryALTinterwordspacing
S.-H. Lee, J.-W. Hwang, M.-H. Song, and H.-M. Park, ``A method based on dual
  cross-modal attention and parameter sharing for polyphonic sound event
  localization and detection,'' \emph{Applied Sciences}, vol.~12, no.~10, 2022.
  [Online]. Available: \url{https://www.mdpi.com/2076-3417/12/10/5075}
\BIBentrySTDinterwordspacing

\bibitem{Shimada2022}
K.~Shimada, Y.~Koyama, S.~Takahashi, N.~Takahashi, E.~Tsunoo, and Y.~Mitsufuji,
  ``Multi-accdoa: Localizing and detecting overlapping sounds from the same
  class with auxiliary duplicating permutation invariant training,'' in
  \emph{IEEE International Conference on Acoustics, Speech and Signal
  Processing (ICASSP)}, Singapore, Singapore, May 2022.

\bibitem{Shimada_SONY_task3_report}
K.~Shimada, N.~Takahashi, Y.~Koyama, S.~Takahashi, E.~Tsunoo, M.~Takahashi, and
  Y.~Mitsufuji, ``Ensemble of accdoa- and einv2-based systems with d3nets and
  impulse response simulation for sound event localization and detection,''
  DCASE2021 Challenge, Tech. Rep., November 2021.

\bibitem{Zheng2021}
X.~Zheng, H.~Chen, and Y.~Song, ``Zheng ustc team’s submission for dcase2021
  task4 – semi-supervised sound event detection,'' DCASE2021 Challenge, Tech.
  Rep., June 2021.

\bibitem{nguyen2022salsa}
T.~N.~T. Nguyen, K.~N. Watcharasupat, N.~K. Nguyen, D.~L. Jones, and W.-S. Gan,
  ``Salsa: Spatial cue-augmented log-spectrogram features for polyphonic sound
  event localization and detection,'' \emph{IEEE/ACM Transactions on Audio,
  Speech, and Language Processing}, vol.~30, pp. 1749--1762, 2022.

\bibitem{Nguyen2022}
T.~N.~T. Nguyen, D.~L. Jones, K.~N. Watcharasupat, H.~Phan, and W.-S. Gan,
  ``{SALSA-Lite: A fast and effective feature for polyphonic sound event
  localization and detection with microphone arrays},'' in \emph{IEEE
  International Conference on Acoustics, Speech and Signal Processing
  (ICASSP)}, Singapore, Singapore, May 2022.

\bibitem{Xie_UESTC_task3_report}
R.~Xie, C.~Shi, L.~Zhang, Y.~Liu, and H.~Li, ``Ensemble of attention based crnn
  for sound event detection and localization,'' DCASE2022 Challenge, Tech.
  Rep., June 2022.

\bibitem{Ko_KAIST_task3_report}
B.-Y. Ko, H.~Nam, S.-H. Kim, D.~Min, S.-D. Choi, and Y.-H. Park, ``Data
  augmentation and squeeze-and-excitation network on multiple dimension for
  sound event localization and detection in real scenes,'' DCASE2022 Challenge,
  Tech. Rep., June 2022.

\bibitem{Xie_XJU_task3_report}
Y.~Xie, Y.~Hu, Y.~Li, S.~Hou, X.~Zhu, Z.~Chen, L.~Wang, and M.~Ma, ``Sound
  event localization and detection based on crnn using time-frequency attention
  and criss-cross attention,'' DCASE2022 Challenge, Tech. Rep., June 2022.

\bibitem{MFCC}
E.~Cakir, T.~Heittola, H.~Huttunen, and T.~Virtanen, ``Polyphonic sound event
  detection using multi label deep neural networks,'' in \emph{2015
  international joint conference on neural networks (IJCNN)}.\hskip 1em plus
  0.5em minus 0.4em\relax IEEE, 2015, pp. 1--7.

\bibitem{araki2007underdetermined}
S.~Araki, H.~Sawada, R.~Mukai, and S.~Makino, ``Underdetermined blind sparse
  source separation for arbitrarily arranged multiple sensors,'' \emph{Signal
  processing}, vol.~87, no.~8, pp. 1833--1847, 2007.

\bibitem{misra2016cross}
I.~Misra, A.~Shrivastava, A.~Gupta, and M.~Hebert, ``Cross-stitch networks for
  multi-task learning,'' in \emph{Proceedings of the IEEE conference on
  computer vision and pattern recognition}, 2016, pp. 3994--4003.

\bibitem{selective}
X.~Li, W.~Wang, X.~Hu, and J.~Yang, ``Selective kernel networks,'' in
  \emph{Proceedings of the IEEE/CVF conference on computer vision and pattern
  recognition}, 2019, pp. 510--519.

\bibitem{conformer}
Z.~Peng, W.~Huang, S.~Gu, L.~Xie, Y.~Wang, J.~Jiao, and Q.~Ye, ``Conformer:
  Local features coupling global representations for visual recognition,'' in
  \emph{2021 IEEE/CVF International Conference on Computer Vision (ICCV)},
  2021, pp. 357--366.

\bibitem{fourstage}
Q.~Wang, J.~Du, H.-X. Wu, J.~Pan, F.~Ma, and C.-H. Lee, ``A four-stage data
  augmentation approach to resnet-conformer based acoustic modeling for sound
  event localization and detection,'' \emph{IEEE/ACM Transactions on Audio,
  Speech, and Language Processing}, vol.~31, pp. 1251--1264, 2023.

\bibitem{Fmix}
E.~Harris, A.~Marcu, M.~Painter, M.~Niranjan, A.~Pr{\"u}gel-Bennett, and
  J.~Hare, ``Fmix: Enhancing mixed sample data augmentation,'' \emph{arXiv
  preprint arXiv:2002.12047}, 2020.

\bibitem{mixup}
H.~Zhang, M.~Cisse, Y.~N. Dauphin, and D.~Lopez-Paz, ``mixup: Beyond empirical
  risk minimization,'' \emph{arXiv preprint arXiv:1710.09412}, 2017.

\bibitem{cutout}
Z.~Zhong, L.~Zheng, G.~Kang, S.~Li, and Y.~Yang, ``Random erasing data
  augmentation,'' in \emph{Proceedings of the AAAI conference on artificial
  intelligence}, vol.~34, no.~07, 2020, pp. 13\,001--13\,008.

\bibitem{postprocessing}
K.~Shimada, Y.~Koyama, N.~Takahashi, S.~Takahashi, and Y.~Mitsufuji, ``Accdoa:
  Activity-coupled cartesian direction of arrival representation for sound
  event localization and detection,'' in \emph{ICASSP 2021 - 2021 IEEE
  International Conference on Acoustics, Speech and Signal Processing
  (ICASSP)}, 2021, pp. 915--919.

\bibitem{Wu_NKU_task3_report}
S.~Wu, S.~Huang, Z.~Liu, and J.~Liu, ``Mlp-mixer enhanced crnn for sound event
  localization and detection in dcase 2022 task 3,'' DCASE2022 Challenge, Tech.
  Rep., June 2022.

\bibitem{Wang_SJTU_task3_report}
Y.~Wang, Y.~Duan, P.~Wang, Y.~Wang, and W.~Xue, ``Improving low-resource sound
  event localization and detection via active learning with domain
  adaptation,'' DCASE2022 Challenge, Tech. Rep., June 2022.

\bibitem{Kang_KT_task3_report}
S.-I. Kang, M.~Keum, K.~Cho, and Y.~Park, ``Track-wise ensemble of crnn models
  with multi-task adpit for sound event localization and detection,'' DCASE2022
  Challenge, Tech. Rep., June 2022.

\bibitem{politis2021dataset}
\BIBentryALTinterwordspacing
A.~Politis, S.~Adavanne, D.~Krause, A.~Deleforge, P.~Srivastava, and
  T.~Virtanen, ``A dataset of dynamic reverberant sound scenes with directional
  interferers for sound event localization and detection,'' in
  \emph{Proceedings of the 6th Detection and Classification of Acoustic Scenes
  and Events 2021 Workshop (DCASE2021)}, Barcelona, Spain, November 2021, pp.
  125--129. [Online]. Available:
  \url{https://dcase.community/workshop2021/proceedings}
\BIBentrySTDinterwordspacing

\bibitem{politis2020overview}
\BIBentryALTinterwordspacing
A.~Politis, A.~Mesaros, S.~Adavanne, T.~Heittola, and T.~Virtanen, ``Overview
  and evaluation of sound event localization and detection in dcase 2019,''
  \emph{IEEE/ACM Transactions on Audio, Speech, and Language Processing},
  vol.~29, pp. 684--698, 2020. [Online]. Available:
  \url{https://ieeexplore.ieee.org/abstract/document/9306885}
\BIBentrySTDinterwordspacing

\bibitem{kingma2014adam}
D.~P. Kingma and J.~Ba, ``Adam: A method for stochastic optimization,''
  \emph{arXiv preprint arXiv:1412.6980}, 2014.

\bibitem{paszke2017automatic}
A.~Paszke, S.~Gross, S.~Chintala, G.~Chanan, E.~Yang, Z.~DeVito, Z.~Lin,
  A.~Desmaison, L.~Antiga, and A.~Lerer, ``Automatic differentiation in
  pytorch,'' 2017.

\bibitem{park2019specaugment}
\BIBentryALTinterwordspacing
D.~S. Park, W.~Chan, Y.~Zhang, C.-C. Chiu, B.~Zoph, E.~D. Cubuk, and Q.~V. Le,
  ``{SpecAugment: A Simple Data Augmentation Method for Automatic Speech
  Recognition},'' in \emph{Proc. Interspeech 2019}, 2019, pp. 2613--2617.
  [Online]. Available: \url{http://dx.doi.org/10.21437/Interspeech.2019-2680}
\BIBentrySTDinterwordspacing

\bibitem{spatialmixup}
R.~Falcón-Pérez, K.~Shimada, Y.~Koyama, S.~Takahashi, and Y.~Mitsufuji,
  ``Spatial mixup: Directional loudness modification as data augmentation for
  sound event localization and detection,'' in \emph{ICASSP 2022 - 2022 IEEE
  International Conference on Acoustics, Speech and Signal Processing
  (ICASSP)}, 2022, pp. 431--435.

\end{thebibliography}


\begin{thebibliography}{10}
\providecommand{\url}[1]{#1}
\def\UrlFont{\rmfamily}
\providecommand{\newblock}{\relax}
\providecommand{\bibinfo}[2]{#2}
\providecommand\BIBentrySTDinterwordspacing{\spaceskip=0pt\relax}
\providecommand\BIBentryALTinterwordstretchfactor{4}
\providecommand\BIBentryALTinterwordspacing{\spaceskip=\fontdimen2\font plus
\BIBentryALTinterwordstretchfactor\fontdimen3\font minus
  \fontdimen4\font\relax}
\providecommand\BIBforeignlanguage[2]{{%
\expandafter\ifx\csname l@#1\endcsname\relax
\typeout{** WARNING: IEEEtran.bst: No hyphenation pattern has been}%
\typeout{** loaded for the language `#1'. Using the pattern for}%
\typeout{** the default language instead.}%
\else
\language=\csname l@#1\endcsname
\fi
#2}}

\bibitem{cJones2003}
C.~Jones, A.~Smith, and E.~Roberts, ``A sample paper in conference
  proceedings,'' in \emph{Proc. IEEE ICASSP}, vol.~II, 2003, pp. 803--806.

\bibitem{eWilliams1999}
E.~Williams, \emph{Fourier Acoustics: Sound Radiation and Nearfield Acoustic
  Holography}.\hskip 1em plus 0.5em minus 0.4em\relax London, UK: Academic
  Press, 1999.

\bibitem{aSmith2000}
A.~Smith, C.~Jones, and E.~Roberts, ``A sample paper in journals,'' \emph{IEEE
  Trans. Signal Process.}, vol.~62, pp. 291--294, Jan. 2000.

\bibitem{dcase2022web}
\url{http://dcase.community/challenge2022/}.

\bibitem{ieeedjo}
\url{http://www.ieee.org/portal/pages/pubs/confpubcenter/register.html}.

\bibitem{IEEEPDFSpec}
{PDF} specification for {IEEE} {X}plore,
  \url{https://www2.securecms.com/ICASSP2015/papers/PaperFormat/Author-PDF-Guide-V32.pdf}.

\bibitem{dcase2019task3}
``Dcase2019 task3 challenge main page,'' \url{
  https://dcase.community/challenge2019/task-sound-event-localization-and-detection},
  accessed on 1 March 2019.

\bibitem{dcase2020task3}
``Dcase2020 task3 challenge main page,'' \url{
  https://dcase.community/challenge2020/task-sound-event-localization-and-detection},
  accessed on 1 March 2020.

\bibitem{dcase2021task3}
``Dcase2021 task3 challenge main page,'' \url{
  https://dcase.community/challenge2021/task-sound-event-localization-and-detection},
  accessed on 1 March 2021.

\bibitem{dcase2022task3}
``Dcase2022 task3 challenge main page,'' \url{
  https://dcase.community/challenge2022/task-sound-event-localization-and-detection},
  accessed on 1 March 2020.

\bibitem{Adavanne2018_JSTSP}
\BIBentryALTinterwordspacing
S.~Adavanne, A.~Politis, J.~Nikunen, and T.~Virtanen, ``Sound event
  localization and detection of overlapping sources using convolutional
  recurrent neural networks,'' \emph{IEEE Journal of Selected Topics in Signal
  Processing}, vol.~13, no.~1, pp. 34--48, March 2018. [Online]. Available:
  \url{https://ieeexplore.ieee.org/abstract/document/8567942}
\BIBentrySTDinterwordspacing

\bibitem{Kong1904.03476}
Q.~Kong, Y.~Cao, T.~Iqbal, Y.~Xu, W.~Wang, and M.~D. Plumbley, ``Cross-task
  learning for audio tagging, sound event detection and spatial localization:
  Dcase 2019 baseline systems,'' \emph{arXiv preprint arXiv:1904.03476}, 2019.

\bibitem{9413473}
Y.~Cao, T.~Iqbal, Q.~Kong, F.~An, W.~Wang, and M.~D. Plumbley, ``An improved
  event-independent network for polyphonic sound event localization and
  detection,'' in \emph{ICASSP 2021 - 2021 IEEE International Conference on
  Acoustics, Speech and Signal Processing (ICASSP)}, 2021, pp. 885--889.

\bibitem{Shimada2022}
K.~Shimada, Y.~Koyama, S.~Takahashi, N.~Takahashi, E.~Tsunoo, and Y.~Mitsufuji,
  ``Multi-accdoa: Localizing and detecting overlapping sounds from the same
  class with auxiliary duplicating permutation invariant training,'' in
  \emph{IEEE International Conference on Acoustics, Speech and Signal
  Processing (ICASSP)}, Singapore, Singapore, May 2022.

\bibitem{Nguyen2022}
T.~N.~T. Nguyen, D.~L. Jones, K.~N. Watcharasupat, H.~Phan, and W.-S. Gan,
  ``{SALSA-Lite: A fast and effective feature for polyphonic sound event
  localization and detection with microphone arrays},'' in \emph{IEEE
  International Conference on Acoustics, Speech and Signal Processing
  (ICASSP)}, Singapore, Singapore, May 2022.

\bibitem{DK}
T.~Kong, S.~Yin, D.~Zhang, W.~Geng, X.~Wang, D.~Song, J.~Huang, H.~Shi, and
  X.~Wang, ``Dynamic multi-scale convolution for dialect identification,''
  \emph{arXiv preprint arXiv:2108.07787}, 2021.

\bibitem{fonseca2022FSD50K}
E.~Fonseca, X.~Favory, J.~Pons, F.~Font, and X.~Serra, ``{FSD50K}: an open
  dataset of human-labeled sound events,'' \emph{IEEE/ACM Transactions on
  Audio, Speech, and Language Processing}, vol.~30, pp. 829--852, 2022.

\bibitem{politis2020dataset}
A.~Politis, S.~Adavanne, and T.~Virtanen, ``A dataset of reverberant spatial
  sound scenes with moving sources for sound event localization and
  detection,'' \emph{arXiv preprint arXiv:2006.01919}, 2020.

\bibitem{conformer}
Z.~Peng, W.~Huang, S.~Gu, L.~Xie, Y.~Wang, J.~Jiao, and Q.~Ye, ``Conformer:
  Local features coupling global representations for visual recognition,'' in
  \emph{2021 IEEE/CVF International Conference on Computer Vision (ICCV)},
  2021, pp. 357--366.

\bibitem{Fmix}
E.~Harris, A.~Marcu, M.~Painter, M.~Niranjan, A.~Pr{\"u}gel-Bennett, and
  J.~Hare, ``Fmix: Enhancing mixed sample data augmentation,'' \emph{arXiv
  preprint arXiv:2002.12047}, 2020.

\bibitem{mixup}
H.~Zhang, M.~Cisse, Y.~N. Dauphin, and D.~Lopez-Paz, ``mixup: Beyond empirical
  risk minimization,'' \emph{arXiv preprint arXiv:1710.09412}, 2017.

\bibitem{fourstage}
Q.~Wang, J.~Du, H.-X. Wu, J.~Pan, F.~Ma, and C.-H. Lee, ``A four-stage data
  augmentation approach to resnet-conformer based acoustic modeling for sound
  event localization and detection,'' \emph{arXiv preprint arXiv:2101.02919},
  2021.

\bibitem{cutout}
Z.~Zhong, L.~Zheng, G.~Kang, S.~Li, and Y.~Yang, ``Random erasing data
  augmentation,'' in \emph{Proceedings of the AAAI conference on artificial
  intelligence}, vol.~34, no.~07, 2020, pp. 13\,001--13\,008.

\bibitem{postprocessing}
K.~Shimada, Y.~Koyama, N.~Takahashi, S.~Takahashi, and Y.~Mitsufuji, ``Accdoa:
  Activity-coupled cartesian direction of arrival representation for sound
  event localization and detection,'' in \emph{ICASSP 2021 - 2021 IEEE
  International Conference on Acoustics, Speech and Signal Processing
  (ICASSP)}, 2021, pp. 915--919.

\bibitem{politis2020overview}
\BIBentryALTinterwordspacing
A.~Politis, A.~Mesaros, S.~Adavanne, T.~Heittola, and T.~Virtanen, ``Overview
  and evaluation of sound event localization and detection in dcase 2019,''
  \emph{IEEE/ACM Transactions on Audio, Speech, and Language Processing},
  vol.~29, pp. 684--698, 2020. [Online]. Available:
  \url{https://ieeexplore.ieee.org/abstract/document/9306885}
\BIBentrySTDinterwordspacing

\bibitem{accdoa}
K.~Shimada, Y.~Koyama, N.~Takahashi, S.~Takahashi, and Y.~Mitsufuji, ``Accdoa:
  Activity-coupled cartesian direction of arrival representation for sound
  event localization and detection,'' in \emph{ICASSP 2021 - 2021 IEEE
  International Conference on Acoustics, Speech and Signal Processing
  (ICASSP)}, 2021, pp. 915--919.

\bibitem{star}
A.~Politis, K.~Shimada, P.~Sudarsanam, S.~Adavanne, D.~Krause, Y.~Koyama,
  N.~Takahashi, S.~Takahashi, Y.~Mitsufuji, and T.~Virtanen, ``Starss22: A
  dataset of spatial recordings of real scenes with spatiotemporal annotations
  of sound events,'' \emph{arXiv preprint arXiv:2206.01948}, 2022.

\bibitem{politis2021dataset}
\BIBentryALTinterwordspacing
A.~Politis, S.~Adavanne, D.~Krause, A.~Deleforge, P.~Srivastava, and
  T.~Virtanen, ``A dataset of dynamic reverberant sound scenes with directional
  interferers for sound event localization and detection,'' in
  \emph{Proceedings of the 6th Detection and Classification of Acoustic Scenes
  and Events 2021 Workshop (DCASE2021)}, Barcelona, Spain, November 2021, pp.
  125--129. [Online]. Available:
  \url{https://dcase.community/workshop2021/proceedings}
\BIBentrySTDinterwordspacing

\bibitem{kingma2014adam}
D.~P. Kingma and J.~Ba, ``Adam: A method for stochastic optimization,''
  \emph{arXiv preprint arXiv:1412.6980}, 2014.

\bibitem{Du_NERCSLIP_task3_report}
Q.~Wang, L.~Chai, H.~Wu, Z.~Nian, S.~Niu, S.~Zheng, Y.~Wang, L.~Sun, Y.~Fang,
  J.~Pan, J.~Du, and C.-H. Lee, ``The nerc-slip system for sound event
  localization and detection of dcase2022 challenge,'' DCASE2022 Challenge,
  Tech. Rep., June 2022.

\bibitem{Hu_IACAS_task3_report}
J.~Hu, Y.~Cao, M.~Wu, Q.~Kong, F.~Yang, M.~D. Plumbley, and J.~Yang, ``Sound
  event localization and detection for real spatial sound scenes:
  Event-independent network and data augmentation chains,'' DCASE2022
  Challenge, Tech. Rep., June 2022.

\bibitem{Han_KU_task3_report}
J.~S. Kim, H.~J. Park, W.~Shin, and S.~W. Han, ``A robust framework for sound
  event localization and detection on real recordings,'' DCASE2022 Challenge,
  Tech. Rep., June 2022.

\bibitem{Xie_UESTC_task3_report}
R.~Xie, C.~Shi, L.~Zhang, Y.~Liu, and H.~Li, ``Ensemble of attention based crnn
  for sound event detection and localization,'' DCASE2022 Challenge, Tech.
  Rep., June 2022.

\bibitem{Kang_KT_task3_report}
S.-I. Kang, M.~Keum, K.~Cho, and Y.~Park, ``Track-wise ensemble of crnn models
  with multi-task adpit for sound event localization and detection,'' DCASE2022
  Challenge, Tech. Rep., June 2022.

\bibitem{Wang_SJTU_task3_report}
Y.~Wang, Y.~Duan, P.~Wang, Y.~Wang, and W.~Xue, ``Improving low-resource sound
  event localization and detection via active learning with domain
  adaptation,'' DCASE2022 Challenge, Tech. Rep., June 2022.

\bibitem{Wu_NKU_task3_report}
S.~Wu, S.~Huang, Z.~Liu, and J.~Liu, ``Mlp-mixer enhanced crnn for sound event
  localization and detection in dcase 2022 task 3,'' DCASE2022 Challenge, Tech.
  Rep., June 2022.

\bibitem{dcase2022result}
``Dcase2022 task3 results page,''
  \url{https://dcase.community/challenge2022/task-sound-event-localization-and-detection-evaluated-in-
  real-spatial-sound-scenes-results}, accessed on 4 July 2022.

\bibitem{park2019specaugment}
D.~S. Park, W.~Chan, Y.~Zhang, C.-C. Chiu, B.~Zoph, E.~D. Cubuk, and Q.~V. Le,
  ``Specaugment: A simple data augmentation method for automatic speech
  recognition,'' \emph{arXiv preprint arXiv:1904.08779}, 2019.

\bibitem{Adavanne2019_DCASE}
\BIBentryALTinterwordspacing
S.~Adavanne, A.~Politis, and T.~Virtanen, ``A multi-room reverberant dataset
  for sound event localization and detection,'' in \emph{Proceedings of the
  Detection and Classification of Acoustic Scenes and Events 2019 Workshop
  (DCASE2019)}, New York University, NY, USA, October 2019, pp. 10--14.
  [Online]. Available: \url{https://dcase.community/workshop2019/proceedings}
\BIBentrySTDinterwordspacing

\bibitem{gemmeke2017audio}
J.~F. Gemmeke, D.~P. Ellis, D.~Freedman, A.~Jansen, W.~Lawrence, R.~C. Moore,
  M.~Plakal, and M.~Ritter, ``Audio set: An ontology and human-labeled dataset
  for audio events,'' in \emph{2017 IEEE international conference on acoustics,
  speech and signal processing (ICASSP)}.\hskip 1em plus 0.5em minus
  0.4em\relax IEEE, 2017, pp. 776--780.

\bibitem{IVs}
D.~Pavlidi, S.~Delikaris-Manias, V.~Pulkki, and A.~Mouchtaris, ``3d
  localization of multiple sound sources with intensity vector estimates in
  single source zones,'' in \emph{2015 23rd European Signal Processing
  Conference (EUSIPCO)}.\hskip 1em plus 0.5em minus 0.4em\relax IEEE, 2015, pp.
  1556--1560.

\bibitem{MFCC}
E.~Cakir, T.~Heittola, H.~Huttunen, and T.~Virtanen, ``Polyphonic sound event
  detection using multi label deep neural networks,'' in \emph{2015
  international joint conference on neural networks (IJCNN)}.\hskip 1em plus
  0.5em minus 0.4em\relax IEEE, 2015, pp. 1--7.

\bibitem{nguyen2022salsa}
T.~N.~T. Nguyen, K.~N. Watcharasupat, N.~K. Nguyen, D.~L. Jones, and W.-S. Gan,
  ``Salsa: Spatial cue-augmented log-spectrogram features for polyphonic sound
  event localization and detection,'' \emph{IEEE/ACM Transactions on Audio,
  Speech, and Language Processing}, vol.~30, pp. 1749--1762, 2022.

\bibitem{spatialmixup}
R.~Falcón-Pérez, K.~Shimada, Y.~Koyama, S.~Takahashi, and Y.~Mitsufuji,
  ``Spatial mixup: Directional loudness modification as data augmentation for
  sound event localization and detection,'' in \emph{ICASSP 2022 - 2022 IEEE
  International Conference on Acoustics, Speech and Signal Processing
  (ICASSP)}, 2022, pp. 431--435.

\bibitem{misra2016cross}
I.~Misra, A.~Shrivastava, A.~Gupta, and M.~Hebert, ``Cross-stitch networks for
  multi-task learning,'' in \emph{Proceedings of the IEEE conference on
  computer vision and pattern recognition}, 2016, pp. 3994--4003.

\end{thebibliography}


%

\begin{IEEEbiography}
[{\includegraphics[width=1in,height=1.25in,clip,keepaspectratio]{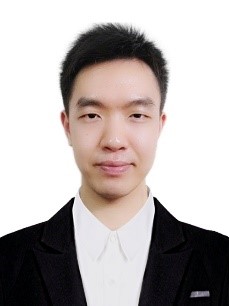}}]{Siwei Huang} received the B.S. degree in Detection
Guidance and Control Technology from School of Marine 
 Science and Technology, Northwestern Polytechnical University,
Xi'an, China in 2021, where he is pursuing an M.S. degree in Information and communication engineering. His research interests include sound event localization and detection, audio tagging and array signal processing.
\end{IEEEbiography}

\begin{IEEEbiography}[{\includegraphics[width=1in,height=1.25in,clip,keepaspectratio]{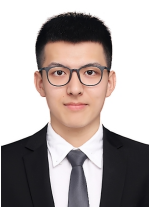}}]{Jisheng Bai}
(Member, IEEE) received the B.S. degree in Detection
Guidance and Control Technology from North University of China in 2017. He received the M.S.
degree in Electronics and Communications Engineering from Northwestern Polytechnical University
in 2020, where he is pursuing the Ph.D. degree. His
research interests are focused on deep learning and
environmental signal processing.
\end{IEEEbiography}

\begin{IEEEbiography}[{\includegraphics[width=1in,height=1.25in,clip,keepaspectratio]{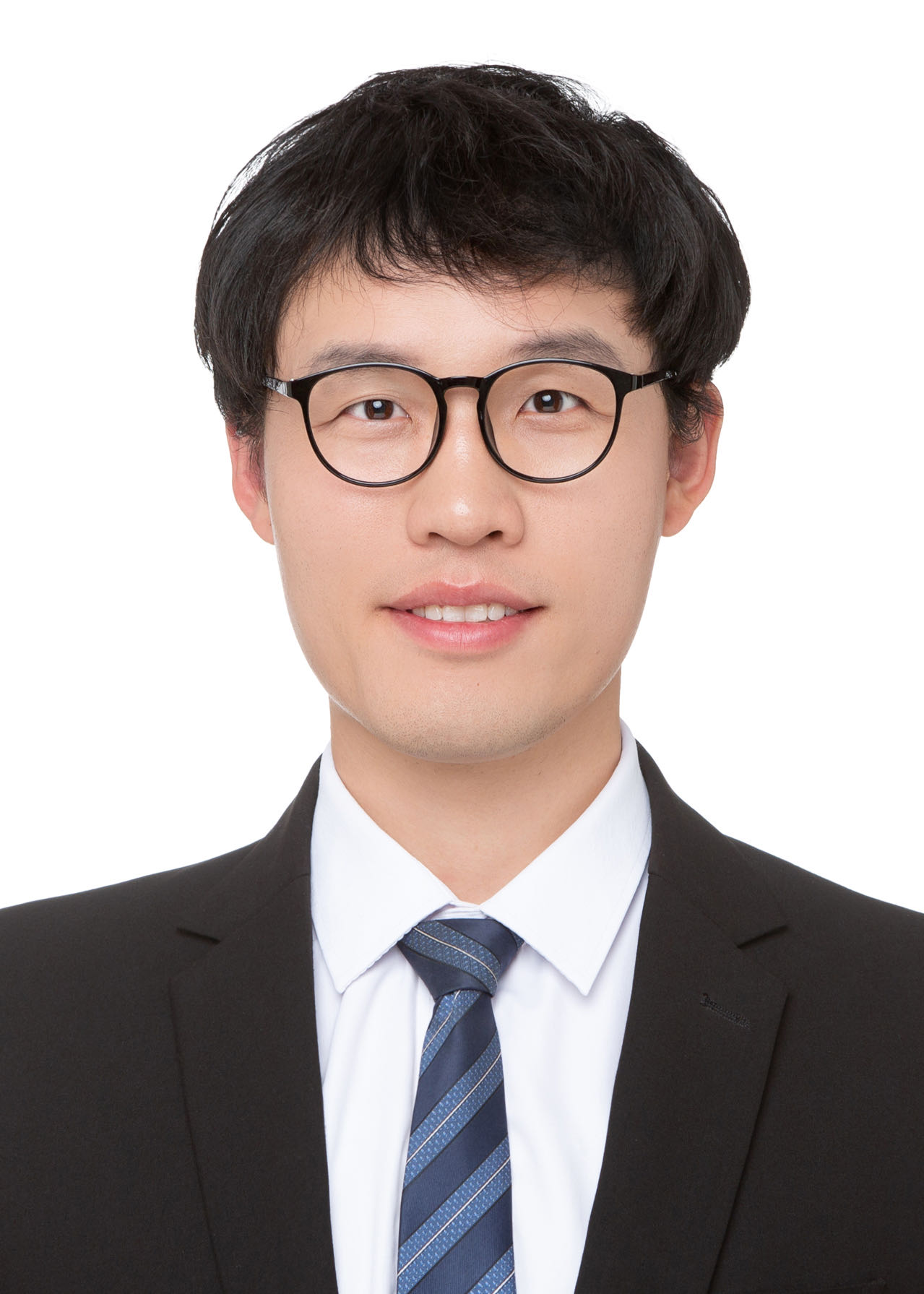}}]{Dongzhe Zhang} received his Master's degree in Electronics and Communication Engineering from Northwestern Polytechnical University in 2022, where he is pursuing the Ph.D. degree. His research interests include source localization and deep learning, especially distributed microphone arrays.

\end{IEEEbiography}

\begin{IEEEbiography}[{\includegraphics[width=1in,height=1.25in,clip,keepaspectratio]{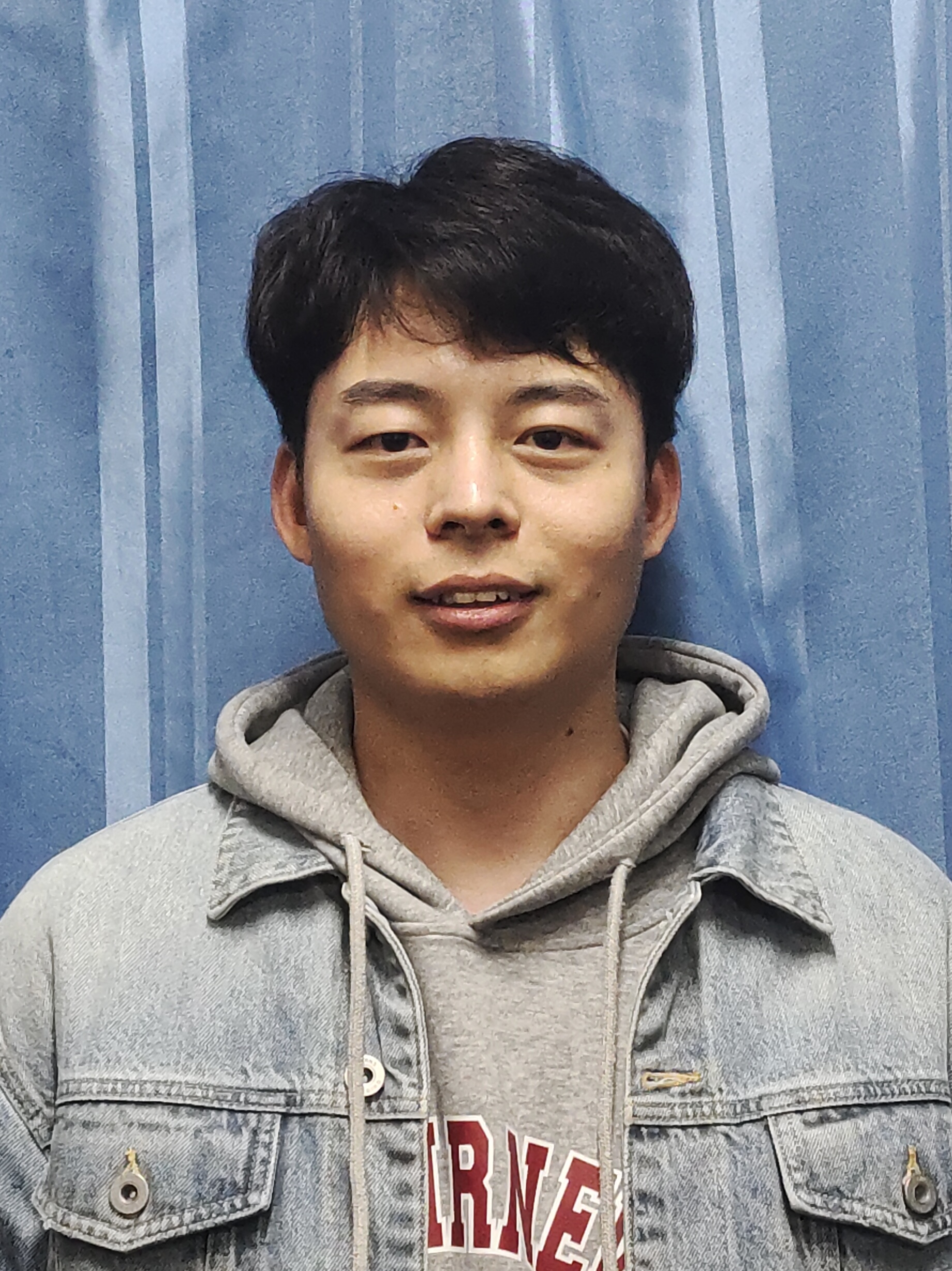}}]{Yafei Jia} received the B.S. degree in Electronic Information Science and Technology from Yantai University in 2020. He received the M.S.degree in Electronics and Communications Engineering from Northwestern Polytechnical University in 2023, where he is pursuing the Ph.D. degree. His research interests are focused on deep learning and abnormal sound detection.

\end{IEEEbiography}


\begin{IEEEbiography}[{\includegraphics[width=1in,height=1.25in,clip,keepaspectratio]{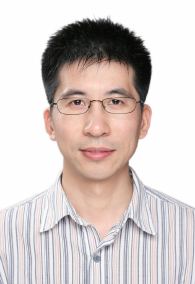}}]{Jianfeng Chen} (Fellow, IEEE) 
received the Ph.D. degree in 1999,
from Northwestern Polytechnical University. From
1999 to 2001, he was a Research Fellow in School of
EEE, Nanyang Technological University, Singapore.
During 2001-2003, he was with Center for Signal
Processing, NSTB, Singapore, as a Research Scientist. From 2003 to 2007, he was a Research Scientist in Institute for Infocomm Research, Singapore.
Since 2007, he has worked in School of Marine Science and Technology as a
professor at Northwestern Polytechnical University,
Xi'an, China. His main research interests include
autonomous underwater vehicle design and application, array processing,
acoustic signal processing, target detection and localization.

\end{IEEEbiography}




\end{document}